\documentclass[11pt, a4paper]{article}
\usepackage{styleBuding}
\usepackage{bm}
\usepackage{listings}
\usepackage[usenames,dvipsnames,svgnames,table]{xcolor}
\usepackage{amsmath}
\usepackage{amssymb}
\usepackage{graphicx}
\usepackage{slashed}
\usepackage{soul}
\usepackage{multirow}
\usepackage{subfigure}
\usepackage{siunitx} % for physics units
\usepackage[utf8]{inputenc}

\pdfoutput=1

\definecolor{back}{HTML}{F8F8F8}

\usepackage{epstopdf} % convert any.eps graphics to.pdf compiling with pdfLaTeX

\begin{document}
%\maketitle
%\newpage

\title{Status of $\mathbb{Z}_3$-NMSSM featuring a light bino-dominated LSP and a light singlet-like scalar under the LZ experiment}

\author{Haijing Zhou\footnote[1]{Email: zhouhaijing0622@163.com },}
\author{Guangning Ban}

\affiliation{School of Physics, Henan Normal University, Henan Xinxiang 453007, China}

\abstract{In the presence of a light singlet-like scalar, the bino-dominated dark matter (DM) candidate in the $\mathbb{Z}_3$-symmetric next-to-minimal supersymmetric standard model ($\mathbb{Z}_3$-NMSSM) exhibits notable deviations from its counterpart in the minimal supersymmetric standard model (MSSM), both in terms of its inherent properties and the mechanisms determining its relic abundance and detection prospects. Motivated by recent progress in experimental particle physics, this study systematically investigates the implications for the \( \mathbb{Z}_3 \)-NMSSM framework featuring a light bino-dominated DM particle and a light singlet-like scalar, ensuring theoretical consistency with empirical observations. Of particular significance are the latest results from the LUX-ZEPLIN (LZ) direct detection experiment, supersymmetry (SUSY) searches at the Large Hadron Collider (LHC), and precision measurements of the Muon g-2 anomaly at Fermilab, which collectively impose complementary constraints on the model's viable parameter space. A comprehensive parameter scan was conducted using the MultiNest algorithm, incorporating constraints from LZ-2022 data, LHC Higgs analyses, Muon g-2 measurements, and B-physics observables. The analysis reveals that current experimental limits---particularly those on spin-independent (SI) and spin-dependent (SD) DM-nucleon scattering cross-sections and LHC constraints on electroweakinos---severely restrict the model. Nevertheless, the framework remains capable of naturally accommodating the observed Z boson and standard model-like Higgs boson masses, accounting for the Muon g-2 anomaly, and inducing sizable corrections to the W boson mass. These results are distinctive to the NMSSM and emerge from the interplay of bino-dominated DM and singlino components, with essential contributions from higgsino.}

'\maketitle

\section{\label{Introduction}Introduction}
The discovery of the Higgs boson in 2012 by the ATLAS and CMS experiments at the large hadron collider (LHC) \cite{Aad:2012tfa,Chatrchyan:2012ufa} validated the Higgs mechanism as the origin of subatomic particle masses. In the decade since this discovery, precision measurements of the Higgs boson's production cross-sections, decay channels, and couplings have been conducted by both collaborations, showing agreement with standard model (SM) predictions within approximately $2\sigma$ confidence levels \cite{ATLAS:2022vkf,CMS:2022dwd}. Nevertheless, a range of theoretical and experimental indications---including the hierarchy problem, absence of gauge coupling unification, non-zero neutrino masses, and the matter-antimatter asymmetry---suggest that SM is an effective low-energy approximation of a more fundamental theory yet to be uncovered.

To address these limitations, various theories beyond the SM have been proposed, many of which extend the scalar sector while retaining the 125 GeV scalar as a key component. Among the most studied are supersymmetry (SUSY) models that conserve R-parity, notably the minimal supersymmetric standard model (MSSM)~\cite{Haber:1984rc,Gunion:1984yn} and the next-to-minimal supersymmetric standard model (NMSSM)~\cite{Ellwanger:2009dp,Haber:1986gz,Maniatis:2009re,Cheung:2010ba}. These frameworks offer an elegant resolution to the hierarchy problem by incorporating quantum corrections from superpartners to stabilize the Higgs mass. Furthermore, R-parity conservation guarantees the stability of the lightest neutralino when it is the lightest supersymmetric particle (LSP), positioning it as a promising dark matter (DM) candidate. While the MSSM stands out for its minimalistic structure and theoretical appeal, it faces persistent challenges, such as the ``$\mu$-problem''~\cite{Kim:1983dt} and the ``little hierarchy problem''~\cite{BasteroGil:2000bw}, both of which have become more pronounced in light of results from LHC Run I. A key tension arises from the unexpectedly large mass of the discovered Higgs boson $m_h \simeq 125~{\rm GeV}$~\cite{Aad:2015zhl,Hall:2011aa,Ellwanger:2011aa, Gunion:2012zd,King:2012tr, Kang:2012sy,King:2012is,Cao:2012fz,Vasquez:2012hn}. By contrast, the NMSSM addresses the $\mu$-problem by introducing a
gauge singlet chiral superfield ($\hat{S}$), extending the MSSM superpotential to
\begin{eqnarray}{\label{Superpotential}}
W=\lambda SH_{u}H_{d}+\frac{\kappa}{3}S^{3}.
\end{eqnarray}
where the effective $\mu$-term arises dynamically as $\mu_{\mathrm{eff}} = \lambda v_s $ after electroweak symmetry breaking. Here, $v_s$ represents the vacuum expectation value (VEV) of $\hat{S}$, and the magnitude of $\mu_{\mathrm{eff}}$ is naturally stabilized at the electroweak scale via the dimensionless Yukawa couplings $\lambda$ and $\kappa$~\cite{Ellwanger:2009dp,Maniatis:2009re}. In the decoupling limit ($ \lambda, \kappa \rightarrow 0 $ and $ v_s \rightarrow \infty $)~\cite{Ellwanger:2009dp}, the NMSSM phenomenologically reduces to an MSSM-like framework.

The Higgs sector of the NMSSM includes two additional singlet-like scalar states---one CP-even $( h_s ) $ and CP-odd $( a_s)$---beyond those present in the MSSM. Moreover, the MSSM neutralino sector is extended through the introduction of a singlet Majorana fermion, commonly referred to as the singlino. As in the MSSM, the lightest neutralino in the NMSSM remains a viable cold DM candidate, typically in the form of a weakly interacting massive particle (WIMP)~\footnote{As a WIMP candidate, DM may be detected through its scattering interactions with nucleons. In the non-relativistic regime, only two primary interaction types are relevant~\cite{Jungman:1995df}: spin-dependent (SD), wherein the WIMP couples to the nucleon's spin, and spin-independent (SI), wherein it couples to the nucleon's mass. Theoretically, WIMPs are expected to interact with SM particles via weak interactions, yielding SI and SD scattering cross-sections on the order of $10^{-45}$ and $10^{-39}\ {\rm cm}^2$, respectively~\cite{Baum:2017enm}. This has prompted extensive experimental efforts for detection, encompassing direct~\cite{XENON:2018voc,PandaX-II:2020oim,PandaX-II:2017hlx,LZ:2022lsv,LZ:2024zvo}, indirect~\cite{PAMELA:2010kea,Fermi-LAT:2011baq,AMS01:2007rrn}, and collider-based approaches~\cite{Goodman:2010yf,Fox:2011pm}. Notably, recent results from direct detection experiments such as LUX-ZEPLIN (LZ) have placed the most stringent upper bounds to date on SI and SD DM-nucleon interactions, limiting them to below $10^{-47}$ and $10^{-42}\ {\rm cm}^2$, respectively~\cite{LZ:2022lsv,LZ:2024zvo}, with no definitive detection observed. These null results suggest that DM-nucleon interactions if they exist, are exceedingly weak.}. The physical masses and compositions of the neutralinos are determined by the mixing between various gauge and singlet components, significantly affecting both DM phenomenology and collider signatures.

Given the increasingly stringent upper limits on SI and SD DM-nucleon scattering cross sections, the DM candidate in the MSSM is typically bino-dominated and achieves the observed relic abundance via co-annihilation with wino-like electroweakinos~\cite{Planck:2018vyg}. In this scenario, the SD scattering cross section, $\sigma^{SD}$, depends solely on the higgsino mass parameter $\mu$ and scales as $1/\mu^4$, leading to significant suppression for large $\mu$. Constraints from the LZ-2022 experiment alone impose a lower bound on $\mu$, requiring $\mu \gtrsim 380$ GeV ~\cite{He:2023lgi,Li:2023hey}. This bound may be further increased by several tens of GeV when radiative corrections to the scattering processes are accounted for~\cite{Bisal:2023fgb,Bisal:2023iip}. Although such large $\mu$ values can naturally arise through the Giudice-Masiero mechanism within gravity-mediated SUSY breaking frameworks~\cite{Giudice:1988yz}, they simultaneously exacerbate the fine-tuning problem associated with accurately predicting the mass of the Z boson. This issue becomes more pronounced when considering the implications of the LHC Higgs boson discovery, along with the absence of signals for DM or supersymmetric particles, as the theory has evolved from the ultraviolet scale down to the electroweak scale~\cite{Arvanitaki:2013yja,Evans:2013jna,Baer:2014ica}.

Due to the distinct configurations of the electroweakinos (EWinos) and Higgs sectors, the LSP in the NMSSM can exhibit properties that differ significantly from those in the MSSM, both in its intrinsic nature and in the mechanisms responsible for its relic abundance and detection prospects. In the NMSSM, both the bino-dominated and singlino-dominated neutralinos emerge as potential DM candidates. In particular, singlino-dominated DM has received extensive attention since the formulation of the NMSSM ~\cite{Cao:2016nix,Ellwanger:2016sur,Xiang:2016ndq,Beskidt:2017xsd,Cao:2018rix,Ellwanger:2018zxt,Domingo:2018ykx,Abdallah:2019znp,Baum:2019uzg,vanBeekveld:2019tqp,
Cao:2019qng,Guchait:2020wqn,Zhou:2021pit,Das:2012rr,Ellwanger:2014hia,Chatterjee:2022pxf,Cao:2022htd,Cao:2023juc,Roy:2024yoh,Bisal:2023mgz}. The phenomenological behavior of singlino-dominated DM is predominantly governed by four parameters: $\lambda$, $\tan\beta$, $m_{\tilde{\chi}_1^0}$, and $\mu_{\rm eff}$~\cite{Zhou:2021pit}. In the limit where the singlet Higgs bosons are sufficiently heavy, both the SI and SD scattering cross sections of DM with nucleons scale as $\lambda^4$. Consequently, data from the LZ experiment typically constrain $\lambda \lesssim 0.1$, leading to two primary mechanisms consistent with the observed relic density ~\cite{Cao:2018rix}: (1)co-annihilation with higgsino-dominated EWinos, or (2) resonant annihilations mediated by singlet-dominated CP-even or CP-odd Higgs bosons. The co-annihilation scenario is only viable within a narrow region of parameter space characterized by $2|\kappa|/\lambda \simeq 1$, $\lambda < 0.1$, and $\mu<400$ GeV~\cite{Zhou:2021pit}, while the resonant annihilation scenario requires $2|m_{\tilde{\chi}_1^0}|$ to be close to the relevant scalar mass. Taken together, these DM constraints impose stringent restrictions on the allowed NMSSM parameter space.

In contrast, systematic investigations focused on bino-dominated DM in the NMSSM remain relatively limited. This is likely due to the absence of direct tree-level interactions between the bino and the additional singlet states, rendering the corresponding configurations largely analogous to those in the MSSM involving bino, higgsinos, and wino~\cite{Beskidt:2017xsd}. However, the phenomenology becomes distinctly NMSSM-specific in scenarios featuring a light singlet-like scalar---typically accompanied by a relatively light singlino-like neutralino~\cite{Ellwanger:2014hia,Ellwanger:2016sur,Ellwanger:2018zxt,Abdallah:2019znp,Guchait:2020wqn,Belanger:2005kh, Gunion:2005rw,Abdallah:2020yag}. Notably, such effects become prominent when the soft singlino mass, defined as $m_{\tilde{S}} \equiv \frac{2 \kappa}{\lambda} \mu_{\mathrm{eff}}$, is comparable in magnitude to both $M_1$ and $\mu_{\mathrm{eff}}$. While singlino-higgsino mixing is directly governed by the coupling ~`$\lambda$', the bino does not couple to the singlino at tree level. Consequently, mixing between the bino and singlino arises only indirectly via the higgsino portal, although this mixing can be enhanced in regions where $|M_1| \lesssim |m_{\tilde{S}}| \sim |\mu_{\mathrm{eff}}|$. Furthermore, the relative hierarchy between $ |m_{\tilde{S}}| $ and $ |\mu_{\mathrm{eff}}| $ determines the nature of the NLSP: singlino-like when \( |m_{\tilde{S}}| < |\mu_{\mathrm{eff}}| \), or higgsino-like for \( |m_{\tilde{S}}| > |\mu_{\mathrm{eff}}| \). In the latter case, the neutralino sector simplifies to a bino-higgsino system analogous to that of the MSSM when the singlino also gets decoupled.

The viability of a highly bino-like DM candidate with a mass below 200 GeV---particularly below 100 GeV---within the $\mathbb{Z}_3$-NMSSM  has been thoroughly examined in Ref.~\cite{Abdallah:2020yag}. While such a scenario is strongly constrained in the MSSM due to stringent bounds from collider and cosmological data, it remains viable in the NMSSM. This is largely attributable to the appearance of new 'blind spots' in the SI and SD DM-nucleon scattering amplitudes, which allow the scenario to evade current direct detection limits ~\cite{Abdallah:2020yag}.
Achieving this viability necessitates a precise turning of the bino-higgsino in the lightest neutralino. Such tuning can be realized by appropriately adjusting the value of$\mu_{\text{eff}}$: it must be increased to suppress the scattering cross sections sufficiently, but not to the extent that it significantly reduces the DM annihilation rate, which would otherwise result in an excessive relic density.
The prospects for observing signals from relatively light EWinos and singlet-like scalar states within the $\mathbb{Z}_3$-NMSSM, particularly in top squark decay channels at the LHC, have been explored in Ref.~\cite{Datta:2022bvg}. The spectra discussed in Refs.~\cite{Abdallah:2020yag,Datta:2022bvg} not only provide phenomenologically interesting configurations but are also theoretically motivated, as they can facilitate a strong first-order electroweak phase transition in the early Universe \cite{Athron:2019teq}. Such a phase transition is a necessary condition for successful electroweak baryogenesis, which could account for the observed baryon asymmetry of the Universe, and may also give rise to a stochastic background of gravitational waves potentially detectable by future experiments \cite{Athron:2023xlk}. Additionally, there has been sustained interest in developing efficient search strategies for light scalar states, independent of the identity of the LSP ~\cite{Dermisek:2005ar,Dermisek:2006wr,Cerdeno:2013cz,Christensen:2013dra, Cerdeno:2013qta, Cao:2013gba, King:2014xwa,Dutta:2014hma, Bomark:2015fga, Ellwanger:2015uaz, Conte:2016zjp, Guchait:2016pes,Baum:2017gbj, Guchait:2017ztk, Ellwanger:2017skc, Barducci:2019xkq,Carena:2022yvx,Gershtein:2020mwi,Spira:1997dg,Winkler:2018qyg}. Correspondingly, LHC searches targeting these states constitute a focused and active experimental program
\cite{CMS:2017dmg,CMS:2018qvj,Aaboud:2018fvk, Aaboud:2018gmx, Aaboud:2018iil, Aaboud:2018esj, Aad:2020rtv, Sirunyan:2018mbx, Sirunyan:2018mot,CMS:2018nsh, CMS:2019spf,CMS:2020ffa,Sirunyan:2020eum,CMS:2012qms,CMS:2015nay,CMS:2017hea,CMS:2018zvv,CMS:2018jid,CMS:2021pcy,CMS:2021bvh,CMS:2024ibt,CMS:2023mep,CMS:2022xxa,CMS:2022fxg,CMS:2023ryd,
ATLAS:2015unc,ATLAS:2018coo,ATLAS:2018jnf,ATLAS:2018pvw,ATLAS:2020ahi,ATLAS:2021ldb,ATLAS:2021hbr,ATLAS:2024zoq,ATLAS:2024vpj,ATLAS:2024itc,Alves:2024ykf}\footnote{The current non-observation of additional Higgs bosons at the LHC imposes significant constraints: these states must either possess masses above $2m_t\sim350~{\rm GeV}$, or predominantly consist of singlet components, thereby suppressing their production cross sections. The NMSSM permits exotic decay modes of the SM-like Higgs boson through light singlet scalars, pseudoscalars, or fermionic states. Recent results from ATLAS and CMS have not excluded non-SM decay channels, instead placing upper limits on their branching ratios (BRs)---12\% for ATLAS~\cite{ATLAS:2022vkf} and 16\% for CMS~\cite{CMS:2022dwd}. These bounds are expected to improve to the 5\%--10\% level via indirect constraints from Higgs coupling measurements~\cite{Liss:2013hbb,CMS:2013xfa}. Consequently, precision Higgs measurements at the LHC remain a powerful tool for testing BSM scenarios that predict such exotic decay modes.}.

In this work, we conduct a detailed investigation of the current status of the $\mathbb{Z}_3$-NMSSM featuring a bino-dominated DM candidate, with a bino fraction exceeding 50\%, and  a light singlet-like scalar. The analysis is performed in light of both theoretical considerations and current experimental constraints, with particular emphasis on the latest results from the LZ direct detection experiment. To realize an additional light scalar state, we consider a Higgs sector configuration in which the next-to-lightest $CP$-even Higgs boson corresponds to the observed SM-like Higgs. In this setup, the lightest $CP$-even (odd) scalar is predominantly singlet-like.
The remainder of this paper is structured as follows. In Section~\ref{sec:model}, we present a concise overview of the key features of the $\mathbb{Z}_3$-NMSSM, with a particular emphasis on the Higgs and neutralino sectors. We also provide analytical expressions relevant to DM annihilation and DM-nucleon scattering cross sections. Section~\ref{sec:NR} details our scanning strategy and discusses the characteristics and viability of the model's parameter space in light of current particle physics constraints. Finally, Section~\ref{sec:sum} summarizes the main conclusions of this study.

\section{\label{sec:model} Next-to-minimal supersymmetric standard model}

\subsection{Fundamental NMSSM properties}

The NMSSM represents the simplest extension of the MSSM through the addition of a gauge-singlet Higgs superfield, $\hat{S}$. The corresponding superpotential is given by~\cite{Maniatis:2009re,Cheung:2010ba,Ellwanger:2009dp}:
\begin{align}
\label{eq:superpotential}
 W_\mathrm{NMSSM}=W_\mathrm{MSSM} + \lambda \hat{S} \hat{H_u} \hat{H_d} + \frac{1}{3} \kappa \hat{S}^3,
\end{align}
where $W_\mathrm{MSSM}$ denotes MSSM superpotential excluding the $\mu$-term, $\lambda$ and $\kappa$ are dimensionless parameters, and $\hat{H}_u$, $\hat{H}_d$ are the  Higgs superfields. This superpotential is the most general form consistent with R-parity conservation and a $Z_3$ discrete symmetry, given the specified field content.

Assuming CP conservation, the Higgs sector of the $\mathbb{Z}_3$-NMSSM is characterized at the tree level by six parameters~\cite{Ellwanger:2009dp,Cheung:2010ba}:
\begin{align}
 \lambda,~ \kappa,~ A_\lambda,~ A_\kappa,~ \mu_{eff},~ \tan\beta,
 \label{eq:six}
\end{align}
where $A_\lambda$ and $A_\kappa$ are soft trilinear coefficients, as defined in Eq. (2.5) of Ref.~\cite{Ellwanger:2009dp}.
In the CP-even scalar basis defined as $H_{\rm SM} \equiv \sin{\beta} Re[H_{u}^0] + \cos{\beta} Re[H_{d}^0], H_{\rm NSM} \equiv \cos{\beta} Re[H_{u}^0] - \sin{\beta} Re[H_{d}^0]$, and $H_{\rm S} \equiv Re[S]$, and in the CP-odd basis defined as $A_{\rm NSM} \equiv \cos{\beta} Im[H_{u}^0] + \sin{\beta} Im[H_{d}^0]$, and $A_{\rm S}\equiv Im[S])$\footnote{$H_u^0$, and $H_d^0$ denote the neutral components of the Higgs doublets $H_u$ and $H_d$, respectively}, the three CP-even Higgs mass eigenstates $h_i = \{ h, H, h_s \}$ and two CP-odd mass eigenstates $a_i = \{a_H, a_s\}$ can be expressed as:
\begin{eqnarray}\label{eq:hi}
h_i &=& V_{h_i}^{\rm SM} H_{\rm SM}+V_{h_i}^{\rm NSM} H_{\rm NSM}+V_{h_i}^{\rm S} H_{\rm S}, \nonumber \\
a_i &=& V_{a_i}^{' \rm NSM} A_{\rm NSM} + V_{a_i}^{'\rm S} A_S,
\end{eqnarray}
where $V$ and $V'$ denote the unitary matrices that diagonalize the CP-even and CP-odd Higgs mass-squared matrices, respectively. In this work, we define the physical Higgs state with the largest $H_{\rm SM}$ component as $h$, and refer to it as the SM-like Higgs boson. The Higgs states with dominant $H_{\rm NSM}$ and $H_{\rm S}$ components are labeled as $H$ and $h_{s}$, respectively. For convenience, we denote the CP-even mass eigenstates by $h_1$, $h_2$, and $h_3$, ordered such that $m_{h_1} < m_{h_2} < m_{h_3}$. Current LHC measurements constrain the couplings of the observed Higgs boson to within approximately $10\%$ of their SM values~\cite{Aad:2019mbh,Sirunyan:2018koj}, implying that $\sqrt{\left (V_{h}^{\rm NSM} \right )^2 + \left ( V_{h}^{\rm S} \right )^2} \lesssim 0.1$ and $|V_{h}^{\rm SM}| \sim 1$. Moreover, in the limit of heavy-charged Higgs bosons, the following approximations hold~\cite{Baum:2017enm}:
\begin{eqnarray}
V_H^{\rm SM} &\approx V_{h_s}^{\rm SM} \sim 0 ~, V_H^{\rm NSM} & \approx V_{h_s}^S \approx \left[1 +\left( \frac{V_{h_s}^{\rm NSM}}{V_{h_s}^S} \right)^2 \right]^{-1/2} \sim 1.
\label{Approximations}
\end{eqnarray}
The singlet masses are independent and may take small values, as they are weakly constrained by current experimental data.

In the $Z_3$-NMSSM, mixtures of bino ($\tilde{B}^0$), wino ($\tilde{W}^0$), higgsino ($\tilde{H}_{d,u}^0$), and
singlino ($\tilde{S}^0$) fields form neutralinos. Assuming a basis of $\psi^0 = (-i \tilde{B}, - i \tilde{W}^0, \tilde{H}_{d}^0, \tilde{H}_{u}^0,\tilde{S})$, the neutralino mass matrix is given by~\cite{Ellwanger:2009dp}:
\begin{align}\label{eq:massmatrix}
{\cal M_{\rm neut}} = \left(
\begin{array}{ccccc}
M_1 & 0 & -m_{Z}c_{\beta}s_{W} & m_{Z}s_{\beta}s_{W} & 0 \\
 & M_2 & m_{Z}c_{\beta}c_{W} & -m_{Z}s_{\beta}c_{W} &0 \\
& & 0 & -\mu_{eff} & -\lambda v s_{\beta}\\
& & & 0 & -\lambda v c_{\beta}\\
& & & & \frac{2 \kappa}{\lambda} \mu_{eff}
\end{array}
\right),
\end{align}
where $M_{1},\ M_{2}$, $\mu_{eff}$ and $m_{\tilde{S}} \equiv \frac{2 \kappa}{\lambda} \mu_{eff} $ are the soft SUSY-breaking mass parameters for the bino, wino, higgsinos, and singlino, respectively;
 $m_{\rm Z}$ is the $Z$-boson mass; $\theta_{\rm w}$ is the Weinberg angle ($c_{W} \equiv \cos\theta_{W}$ and $s_{W} \equiv \sin\theta_{W}$);
$\tan\beta\equiv s_\beta/c_\beta=v_{u}/v_{d}$ is the ratio of the VEVs of the two Higgs doublets ($c_\beta \equiv \cos\beta$ and $s_\beta \equiv \sin\beta$) and $v^2=v_{u}^2+v_{d}^2=(174{~\rm GeV})^2$.
Diagonalizing  $M_{\rm neut}$ with a $5\times5$ unitary matrix N yields the masses of the physical neutralino states $\tilde\chi^0_i$ (ordered by mass):
\begin{eqnarray}
 N^* M_{\rm neut} N^{-1} = {\rm diag} \{ m_{\tilde\chi^0_1}, m_{\tilde\chi_2^0}, m_{\tilde\chi_3^0}, m_{\tilde\chi_4^0}, m_{\tilde\chi_5^0} \} \nonumber
\end{eqnarray}
with
\begin{eqnarray}\label{2}
 \tilde\chi^0_i = N_{i1}\tilde B^0 + N_{i2}\tilde W^0 + N_{i3}\tilde H^0_d + N_{i4}\tilde H^0_u + N_{i5}\tilde S ~~~~(i=1,2,3,4,5), \nonumber
\end{eqnarray}
where $m_{\tilde\chi^0_i}$ is the root to the characteristic equation:
\begin{eqnarray}
\left(x-M_1\right) \left(x-M_2\right)\left[ (x^2-\mu_{eff}^2 )\left(m_{\tilde{S}}-x\right)+\lambda^2 v^2\left(x-s_{2\beta}\mu_{eff}\right)\right]\\ \nonumber
-m_Z^2 \left(x-M_1c_W^2-M_2 s_W^2\right) \left[\left(\mu s_{2\beta }+x\right)\left(m_{\tilde{S}}-x\right)+\lambda^2 v^2\right] =0.
\end{eqnarray}
The eigenvector of $m_{\tilde\chi^0_i}$ is the column vector consisting of $N_{ij}(j=1,2,3,4,5)$, which is given by
\begin{equation}\label{NIJ}
N_i =\frac{1}{\sqrt{C_i}}\left(\begin{array}{c}
(M_2- m_{\tilde\chi^0_i})\left[ (m_{\tilde\chi^0_i}^2-\mu_{eff}^2 )\big(m_{\tilde{S}}-m_{\tilde\chi^0_i}\big)+\lambda^2 v^2\big(m_{\tilde\chi^0_i}-s_{2\beta}\mu_{eff}\big)\right]s_W \\
-(M_1- m_{\tilde\chi^0_i})\left[ (m_{\tilde\chi^0_i}^2-\mu_{eff}^2 )\big(m_{\tilde{S}}-m_{\tilde\chi^0_i}\big)+\lambda^2 v^2\big(m_{\tilde\chi^0_i}-s_{2\beta}\mu_{eff}\big)\right]c_W \\
-(M_2 s^2_W + M_1 c^2_W -m_{\tilde\chi^0_i}) \left[\big(m_{\tilde{S}}-m_{\tilde\chi^0_i}\big)(m_{\tilde\chi^0_i}c_\beta + \mu s_\beta)+\lambda^2 v^2 c_{\beta}\right]m_Z \\
(M_2 s^2_W + M_1 c^2_W -m_{\tilde\chi^0_i}) \left[\big(m_{\tilde{S}}-m_{\tilde\chi^0_i}\big)(m_{\tilde\chi^0_i}s_\beta + \mu c_\beta)+\lambda^2v^2s_{\beta}\right]m_Z\\
(M_2 s^2_W + M_1 c^2_W -m_{\tilde\chi^0_i}) \lambda v \mu_{eff} c_{2\beta} m_Z \\
\end{array}\right).
\end{equation}
The specific form of the normalization factor $C_i$ is given by:
\begin{eqnarray}\label{5}
 C_i &=& \bigg[ \big(m_{\tilde{S}}-m_{\tilde\chi^0_i}\big)
 \big(m_{\tilde\chi^0_i}^2-\mu_{eff}^2 \big)-\lambda^2 v^2\big(s_{2\beta}\mu_{eff}-m_{\tilde\chi^0_i}\big)\bigg]^2  \nonumber \\
 &&\bigg[\big(M_2- m_{\tilde\chi^0_i}\big)^2 s_W^2+\big(M_1- m_{\tilde\chi^0_i}\big)^2 c_W^2 \bigg]
 + m_Z^2\big(M_2 s_W^2 + M_1 c_W^2 - m_{\tilde\chi^0_i}\big)^2\nonumber \\
 &&\Bigg [\big(m_{\tilde{S}}-m_{\tilde\chi^0_i}\big)^2\big(\mu_{eff}^2 + m^2_{\tilde\chi^0_i} + 4\mu_{eff} m_{\tilde\chi^0_i}s_\beta c_\beta \big) \nonumber \\
 &&+2\lambda^2v^2\big(m_{\tilde{S}}-m_{\tilde\chi^0_i}\big)\big(\mu_{eff}s_{2\beta}+m_{\tilde\chi^0_i}\big)
 +\lambda^2v^2\mu_{eff}^2c_{2\beta}^2+\lambda^4v^4 \Bigg ].
\end{eqnarray}
Therefore, the diagonalizing matrix is $N=\{N_1,N_2,N_3,N_4,N_5\}$, where ${i=1,2,3,4,5}$ denotes the $i$-th neutralino.

This work focuses on the lightest neutralino, $\tilde\chi^0_1$, which serves as the DM candidate. $N_{11}^2$, $N_{12}^2$, $N_{13}^2+N_{14}^2$, and $N_{15}^2$
 are the bino, wino, higgsino, and singlino components
in the physical state $\tilde\chi^0_1$, respectively, and satisfy $N_{11}^2+N_{12}^2 +N_{13}^2+N_{14}^2 + N_{15}^2=1$. If $N_{11}^2 > 0.5 $($N_{12}^2>0.5$, or $N_{13}^2+N_{14}^2>0.5$, or $N_{15}^2>0.5$), we classify $\tilde\chi^0_1$ as a bino-, wino-, higgsino-, or singlino-dominant DM, respectively. In the following discussion, we focus on the case of a bino-dominant ${\tilde\chi^0_1}$ (i.e., $m_{\tilde\chi^0_1} \approx M_1 $).

The couplings of DM candidates to the scalar Higgs and the Z-boson play a crucial role in the calculation of both DM-nucleon scattering cross-sections and DM annihilation rates. These interactions are described by the following Lagrangian~\cite{Ellwanger:2009dp}:
\begin{eqnarray}
{\cal{L}}_{\rm NMSSM} \ni i C_{\tilde{\chi}_1^0 \tilde{\chi}_1^0 G^0} G^0 \overline{\tilde{\chi}_1^0} \gamma_5 \tilde{\chi}_1^0 + C_{\tilde{\chi}_1^0 \tilde{\chi}_1^0 h_i} h_i \overline{\tilde{\chi}_1^0} \tilde{\chi}_1^0 + i C_{\tilde{\chi}_1^0 \tilde{\chi}_1^0 a_i} a_i \overline{\tilde{\chi}_1^0} \gamma_5 \tilde{\chi}_1^0+C_{\tilde{\chi}_1^0 \tilde{\chi}_1^0 Z} Z_\mu \overline{\tilde{\chi}_1^0} \gamma^\mu \gamma_5 \tilde{\chi}_1^0, \nonumber
\end{eqnarray}
where the coefficients of these couplings are given by~\cite{Ellwanger:2009dp,Baum:2017enm}
\begin{align}
&C_{\tilde {\chi}^0_1 \tilde {\chi}^0_1 h_i }= V_{h_i}^{\rm SM} \bigg[ \sqrt{2} \lambda N_{15} \left( N_{13} s_\beta + N_{14} c_\beta \right)
+ \left( g_1 N_{11} - g_2 N_{12} \right) \left( N_{13} c_\beta - N_{14} s_\beta \right) \bigg]
\nonumber \\
&+V_{h_i}^{\rm NSM} \bigg[ \sqrt{2} \lambda N_{15} \left( N_{13} c_\beta - N_{14} s_\beta \right)
- \left(g_1 N_{11} - g_2N_{12} \right) \left( N_{13} s_\beta + N_{14} c_\beta \right)\bigg]
\nonumber \\
&+V_{h_i}^{\rm S} \bigg[\sqrt{2} \left( \lambda N_{13} N_{14} - \kappa N_{15} N_{15} \right) \bigg],
%&g_{\tilde {\chi}^0_1 \tilde {\chi}^0_1 H^{\rm SM}} =- \frac{\sqrt{2}m_Z^2 }{ v C_1}\left(M_2 s^2_W + M_1 c^2_W -m_{\tilde\chi^0_i}\right)^2\bullet \Bigg[ \lambda^4 v^4\big(m_{\tilde\chi^0_1}-\mu_{eff}s_{2\beta}\big) \nonumber\\
% &+2\lambda^2 v^2\bigg(\frac{2\kappa}{\lambda}\mu_{eff}-m_{\tilde\chi^0_1}\bigg)\big(m_{\tilde\chi^0_1}^2-\mu_{eff}^2 \big)+\bigg(\frac{2\kappa}{\lambda}\mu_{eff}-m_{\tilde\chi^0_1}\bigg)^2
% \big(m_{\tilde\chi^0_i}^2-\mu_{eff}^2 \big)(m_{\tilde\chi^0_1}+\mu_{eff}s_{2\beta}) \Bigg]
%\nonumber \\
%& g_{\tilde {\chi}^0_1 \tilde {\chi}^0_1 H^{\rm NSM} }=- \frac{\sqrt{2}m_Z^2 \mu_{eff}c_{2\beta}}{ v C_1}\left(M_2 s^2_W + M_1 c^2_W -m_{\tilde\chi^0_1}\right)^2\bullet \nonumber\\
%&\Bigg[ \bigg(\frac{2\kappa}{\lambda}\mu_{eff}-m_{\tilde\chi^0_1}\bigg)^2
% \big(m_{\tilde\chi^0_1}^2-\mu_{eff}^2 \big)+2\lambda^2 v^2m_{\tilde\chi^0_1}\big(m_{\tilde\chi^0_1}-\mu_{eff}s_{2\beta}\big)+\lambda^4 v^4\Bigg],
%\nonumber \\
%&g_{\tilde {\chi}^0_1 \tilde {\chi}^0_1 H^{\rm S}} =- \frac{\sqrt{2}m_Z^2 }{ v C_1}\left(M_2 s^2_W + M_1 c^2_W -m_{\tilde\chi^0_1}\right)^2\bullet \Bigg[\lambda^5 v^5s_\beta c_\beta+ \kappa\lambda^2 v^3\mu_{eff}^2c_{2\beta}^2 \nonumber\\
%&+\lambda^3 v^3\bigg(\frac{2\kappa}{\lambda}\mu_{eff}-m_{\tilde\chi^0_1}\bigg)\big(m_{\tilde\chi^0_1}s_{2\beta}+\mu_{eff} \big) \nonumber\\
%&+\lambda v\bigg(\frac{2\kappa}{\lambda}\mu_{eff}-m_{\tilde\chi^0_1}\bigg)^2
% \Big[(m_{\tilde\chi^0_1}^2+\mu_{eff}^2)s_{\beta}c_{\beta}+m_{\tilde\chi^0_1}\mu_{eff}\Big] \Bigg]
\label{eq0:hichi01chi01}
\\
&C_{\tilde {\chi}^0_1 \tilde {\chi}^0_1 a_i }= -P_{a_i}^{\rm NSM}\bigg[ \sqrt{2} \lambda N_{15} \left( N_{13} c_\beta + N_{14} s_\beta \right) - \left( g_1 N_{11} - g_2 N_{12} \right) \left( N_{13} s_\beta - N_{14} c_\beta \right) \bigg]
\nonumber \\
&+ P_{a_i}^{\rm S} \bigg[\sqrt{2} \left(\lambda N_{13} N_{14} - \kappa N_{15} N_{15} \right)\bigg],
\\
&C_{\tilde{\chi}_1^0 \tilde{\chi}_1^0 Z}= \frac{m_Z}{\sqrt{2} v}\left(N_{13}^2-N_{14}^2\right),
%&C_{\tilde{\chi}_1^0 \tilde{\chi}_1^0 Z}= \frac{m_Z^3 c_{2\beta}}{\sqrt{2} v C_1}\left(M_2 s^2_W + M_1 c^2_W -m_{\tilde\chi^0_i}\right)^2\bullet \nonumber\\
%&\Bigg[ \bigg(\frac{2\kappa}{\lambda}\mu_{eff}-m_{\tilde\chi^0_i}\bigg)^2
% \big(m_{\tilde\chi^0_i}^2-\mu_{eff}^2 \big)+2\lambda^2 v^2m_{\tilde\chi^0_i}\big(m_{\tilde\chi^0_i}-\mu_{eff}s_{2\beta}\big)+\lambda^4 v^4\Bigg],
\label{eq0:zchi10chi10}
\\
&C_{\tilde {\chi}^0_1 \tilde {\chi}^0_1 G^0 } = - \sqrt{2} \lambda N_{15} \left( N_{13} s_\beta - N_{14} c_\beta \right) - \left( g_1 N_{11} - g_2 N_{12} \right) \left( N_{13} c_\beta + N_{14} s_\beta \right).
%&C_{\tilde {\chi}^0_1 \tilde {\chi}^0_1 G^0 } = - \frac{i\sqrt{2}m_Z^2 m_{\tilde\chi^0_1}c_{2\beta}}{ v C_1}\left(M_2 s^2_W + M_1 c^2_W -m_{\tilde\chi^0_i}\right)^2\bullet \nonumber\\
%&\Bigg[ \bigg(\frac{2\kappa}{\lambda}\mu_{eff}-m_{\tilde\chi^0_i}\bigg)^2
% \big(m_{\tilde\chi^0_i}^2-\mu_{eff}^2 \big)+2\lambda^2 v^2m_{\tilde\chi^0_i}\big(m_{\tilde\chi^0_i}-\mu_{eff}s_{2\beta}\big)+\lambda^4 v^4\Bigg],
\label{eq0:G0chi10chi10_S}
\end{align}
It is evident that the interactions critical for DM phenomenology heavily depend on the higgsino and singlino admixtures in the LSP, which is considered bino-dominant in our study. As will be demonstrated later, these admixtures also influence collider phenomenology in a significant manner. Additionally, it is important to note that these admixtures exhibit complex variations across the NMSSM parameter space. According to Eq.(\ref{eq:massmatrix}), the variations with $\tan\beta$ and $\lambda$ are especially significant. Moreover, the signs of the various mass parameters present in Eq.(\ref{eq:massmatrix}) also play a crucial role in determining these admixtures.
%

\iffalse
\begin{align}
C_{\tilde{\chi}_1^0 \tilde{\chi}_1^0 Z}= & \frac{m_Z}{\sqrt{2} v}\left(N_{13}^2-N_{14}^2\right),
\label{eq0:zchi10chi10_S}
\\
C_{\tilde {\chi}^0_1 \tilde {\chi}^0_1 G^0 } = & - \sqrt{2} \lambda N_{15} \left( N_{13} s_\beta - N_{14} c_\beta \right) - \left( g_1 N_{11} - g_2 N_{12} \right) \left( N_{13} c_\beta + N_{14} s_\beta \right),
\label{eq0:G0chi10chi10_S}
\\
C_{\tilde {\chi}^0_1 \tilde {\chi}^0_1 h_i }= &
V_{h_i}^{\rm SM}\Bigg[\sqrt{2} \lambda N_{15} \left( N_{13} s_\beta + N_{14} c_\beta \right) + \left( g_1 N_{11} - g_2 N_{12} \right) \left( N_{13} c_\beta - N_{14} s_\beta \right)\Bigg] \nonumber \\
&+V_{h_i}^{\rm NSM} \Bigg[\sqrt{2} \lambda N_{15} \left( N_{13} c_\beta - N_{14} s_\beta \right) - \left(g_1 N_{11} - g_2
N_{12} \right) \left( N_{13} s_\beta + N_{14} c_\beta \right)\Bigg] \nonumber \\
&+V_{h_i}^{\rm S}\Bigg[ \sqrt{2} \left[ \lambda N_{13} N_{14} - \kappa N_{15} N_{15} \right] \Bigg]\,
\label{eq0:hichi01chi01_S}
\\
C_{\tilde {\chi}^0_1 \tilde {\chi}^0_1 a_i }
= &
-V_{a_i}^{' \rm NSM} \left[ \sqrt{2} \lambda N_{15} \left( N_{13} c_\beta + N_{14} s_\beta \right) - \left( g_1 N_{11} - g_2 N_{12} \right) \left( N_{13} s_\beta - N_{14} c_\beta \right) \right] \nonumber \\
&+ V_{a_i}^{' \rm S}\sqrt{2} \left[ \lambda N_{13} N_{14} - \kappa N_{15} N_{15} \right]. \nonumber \\
\label{eq0:aichi01chi01_S}
\end{align}
\fi

\subsection{\label{DM}Dark matter sector}

When the squark mass $m_{\tilde{q}}$ exceeds $ 2~{\rm TeV}$, the t-channel Z boson exchange diagram becomes the dominant contribution to the SD scattering cross-section at the tree-level. This contribution can be parameterized approximately as~\cite{Pierce:2013rda,Calibbi:2014lga}
\begin{eqnarray}
\label{eq:sigSD}
\sigma_{\tilde{\chi}_1^0-N}^{\rm SD} \simeq C_N \times \left ( \frac{C_{\tilde{\chi}_1^0 \tilde{\chi}_1^0 Z}}{0.01} \right )^2,
\end{eqnarray}
with $N=p(n)$ denoting protons (neutrons), and $C_p \simeq 2.9 \times 10^{-41}~{\rm cm^2} $ ($C_n \simeq 2.3 \times 10^{-41}~{\rm cm^2} $)~\cite{Badziak:2015exr,Badziak:2017uto}, Eq.(\ref{eq0:zchi10chi10}) implies that $C_{\tilde{\chi}_1^0 \tilde{\chi}_1^0 Z} \propto N_{13}^2 - N_{14}^2$~\cite{Abdallah:2020yag}. This interaction influences both the DM annihilation cross section through the Z-funnel channel and the SD scattering cross section. Notably, a destructive interference effect, commonly termed a "blind-spot", occurs when $N_{13}^2\approx N_{14}^2$. Through explicit substitution of $N_{13} $ and $ N_{14} $ defined in Eq. (\ref{NIJ}), it can be determined:

\begin{align}\label{eq1:zchi10chi10}
N^2_{13} - N^2_{14}\approx& \frac{m_Z^2 c_{2\beta}}{C_1}\big(M_2-m_{\tilde\chi^0_1}\big)^2 s_w^4 \mu_{eff}^4 \nonumber\\
&\Bigg[ \bigg(\frac{m_{\tilde{S}}}{\mu_{eff}}-\frac{m_{\tilde\chi^0_1}}{\mu_{eff}}\bigg)^2
 \bigg(\frac{m_{\tilde\chi^0_1}^2}{\mu_{eff}^2}-1 \bigg)+2\bigg(\frac{\lambda v}{\mu_{eff}}\bigg)^2\frac{m_{\tilde\chi^0_1}}{\mu_{eff}}\bigg(\frac{m_{\tilde{S}}}{\mu_{eff}}-\frac{m_{\tilde\chi^0_1}}{\mu_{eff}}\bigg)+\bigg(\frac{\lambda v}{\mu_{eff}}\bigg)^4\Bigg].
\end{align}
A condition for an MSSM-like SD blind spot naturally arises when $\tan\beta=1$, where $\cos2\beta$ vanishes~\cite{Cheung:2012qy}. Additionally, significant suppression of the SD cross section can be achieved when cancellations occur among the terms in Eq. (\ref{eq1:zchi10chi10}), a scenario particularly relevant for a bino-dominated LSP coexisting with light singlino-like states \cite{Cheung:2012qy}. The algebraic structure reveals three key features for a bino-dominated ${\tilde{\chi}^0_1}$: (i) The first term is negative definite, while the last term is positive definite; (ii) The sign of the second term depends on the relative sign between $ m_{\tilde{S}} $ and $ m_{\tilde{\chi}^0_1} $, being positive (negative) for identical (opposite) signs. When the signs coincide, the first term must counterbalance the combined contributions of the latter two terms. Conversely, when the signs are opposite, cancellation requires mutual annihilation between the first two terms and the final term, which is theoretically less natural. Importantly, neither $ m_{\tilde{S}} $ nor $ \mu_{eff} $ can significantly exceed the mass of the bino-like LSP to maintain an overall cancellation in the presence of the first term.

Another relevant quantity in the context is
$N^2_{13} + N^2_{14}$, which measures the total higgsino admixture in the LSP and controls scalar (Higgs boson) couplings to a pair of bino-like LSPs. Using Eq.~(\ref{NIJ}), one finds:
\begin{align}\label{eqn:total-higgsino}
N^2_{13} + N^2_{14}=& \frac{m_Z^2} {C_1}\left(M_2 s^2_W + M_1 c^2_W -m_{\tilde\chi^0_1}\right)^2 \mu_{eff}^4  \nonumber\\
&\Bigg[ \big(\frac{m_{\tilde{S}}}{\mu_{eff}}-\frac{m_{\tilde\chi^0_1}}{\mu_{eff}}\big)^2
 \big(1+\big(\frac{m_{\tilde\chi^0_1}}{\mu_{eff}}\big)^2+\frac{m_{\tilde\chi^0_1}}{\mu_{eff}}s_{2\beta} \big) \nonumber\\
& +2\bigg(\frac{\lambda v}{\mu_{eff}}\bigg)^2(\frac{m_{\tilde\chi^0_1}}{\mu_{eff}}+s_{2\beta})\big(\frac{m_{\tilde{S}}}{\mu_{eff}}-\frac{m_{\tilde\chi^0_1}}{\mu_{eff}}\big)+\bigg(\frac{\lambda v}{\mu_{eff}}\bigg)^4\Bigg].
\end{align}
A closer examination shows that Eqs.~(\ref{eq1:zchi10chi10}) and~(\ref{eqn:total-higgsino}) exhibit symmetric under simultaneous sign flips of each of the parameters:
$\mu_{eff}$, $m_{\tilde\chi^0_1} \, (M_1)$, and $m_{\tilde{S}}$.
%

%%\subsection{Comments on the spin-independent scattering cross-section}
In contrast, the SI scattering cross section in the heavy squark limit is dominated by a $t$-channel exchange of CP-even Higgs bosons $h_i$ and is given by~\cite{Drees1993,Drees1992,Jungman1995,Belanger2008}
\begin{align} \label{eq:SIDD_p}
\sigma_{\tilde {\chi}^0_1-{N}}^{\rm SI} &= \frac{ m_N^2}{2\pi v^2} \left( \frac{m_N m_{\tilde{\chi}_1^0}}{m_N + m_{\tilde{\chi}_1^0}} \right)^2 {\left( \frac{1}{125~\rm GeV} \right)^4}\left|f^{(N)}\right|^2, \nonumber\\
 f^{(N)} &= \sum_{i=1}^3 \left( \frac{125 GeV}{ m_{_{h_i}}} \right)^2 {C_{ \tilde{\chi}_1^0 \tilde{\chi}_1^0 h_{i}}\, C_{{h_{i} N N}}}.
\end{align}
$C_{{h_{i} N N}}$ is given by~\cite{Badziak:2015exr}
%
%\begin{eqnarray}
%\label{eqn:hiNN-newbasis}
%C_{{h_i N N}} &= &
%\Big[ V_{h_i}^{\rm SM} (F^{(N)}_d + F^{(N)}_u) + V_{h_i}^{\rm NSM} \big(\tan\beta F^{(N)}_d - \frac{1}{\tan\beta} F^{(N)}_u\big) \Big] \,.
%\end{eqnarray}
%
%
\begin{align}\label{eqn:ChiNN }
&C_{{h_i N N}} = (V_{h_i}^{\rm SM} - V_{h_i}^{\rm NSM} \tan\beta)F^{(N)}_d + (V_{h_i}^{\rm SM}+ V_{h_i}^{\rm NSM} \frac{1}{\tan\beta})F^{(N)}_u.
\end{align}
We define
\begin{align}
&{\cal{B}}_d= \sum_{i=1}^3 \left( \frac{125 GeV}{ m_{_{h_i}}} \right)^2 C_{ \tilde{\chi}_1^0 \tilde{\chi}_1^0 h_{i}}\,(V_{h_i}^{\rm SM} - V_{h_i}^{\rm NSM} \tan\beta), \label{eqn0:BD}\\
&{\cal{B}}_u=\sum_{i=1}^3 \left( \frac{125 GeV}{ m_{_{h_i}}} \right)^2 C_{ \tilde{\chi}_1^0 \tilde{\chi}_1^0 h_{i}}\,(V_{h_i}^{\rm SM}+V_{h_i}^{\rm NSM} \frac{1}{\tan\beta}),
\label{eqn0:BU}
\end{align}
then, $f^{(N)}$ in Eq.(\ref{eq:SIDD_p}) can be rewritten
\begin{align}\label{eqn:FN}
f^{(N)}= {\cal{B}}_dF^{(N)}_d+{\cal{B}}_uF^{(N)}_u,
 \end{align}
where $m_N$ denotes the nucleon mass, $F^{(N)}_d=f^{(N)}_d+f^{(N)}_s+\frac{2}{27}f^{(N)}_G$ and $F^{(N)}_u=f^{(N)}_u+\frac{4}{27}f^{(N)}_G$, with $f^{(N)}_q =m_N^{-1}\left<N|m_qq\bar{q}|N\right> $ ($q=u,d,s$) representing the normalized light quark contribution to the nucleon mass. Additionally, $f^{(N)}_G=1-\sum_{q=u,d,s}f^{(N)}_q$ accounts for the influence of other heavy quark mass fractions in nucleons~\cite{Drees1993,Drees1992}. For the respective form factors for ${u, d, s}$, we use the default parameters implemented in micrOMEGAs 5.2~\cite{Belanger2008}
\begin{eqnarray}
f^p_{q=u,d,s} = \{0.0153, 0.0191, 0.0447\}; ~~~f^n_{q=u,d,s} = \{0.011, 0.0273, 0.0447\},	
\end{eqnarray}
resulting in $F_u^{(p)} \simeq 0.152, F_u^{(n)} \simeq 0.147$, and $F_d^{(p)}\simeq0.132, F_d^{(n)} \simeq 0.140$.
Therefore, the SI cross sections for DM-proton and DM-neutron scattering are generally nearly equal (i.e., $\sigma^{SI}_{{\tilde {\chi}^0_1}-p} \simeq \sigma^{SI}_{{\tilde {\chi}^0_1}-n}$)~\cite{Riffard:2016oss}. However, when ${\cal{B}}_d$ and ${\cal{B}}_u$ in Eq.(\ref{eqn:FN}) have opposite signs and comparable magnitudes(i.e., a strong cancellation effect ), the SI cross sections for proton and neutron diverge significantly. In such cases, the effective cross-section for coherent scattering of DM with xenon nuclei can be computed via~\cite{Cao:2019aam}
\begin{eqnarray}
\sigma_{\rm eff}^{\rm SI} = 0.169 \sigma^{\rm SI}_p + 0.347 \sigma^{\rm SI}_n + 0.484 \sqrt{ \sigma^{\rm SI}_p \sigma^{\rm SI}_n }, \label{Effective-Cross-Section}
\end{eqnarray}
where the three coefficients on the right side are derived from the average abundances of different xenon isotopes in nature. This equivalence clearly demonstrates that the effective cross-section reduces to $\sigma^{\rm SI}_p$ when $\sigma_p^{\rm SI} = \sigma_n^{\rm SI}$.

Current LHC experimental constraints require the non-SM doublet Higgs boson $H$ to have masses greater than several hundred GeV. Under these conditions, its contribution to the SI cross-section is exponentially suppressed by $ (125~\mathrm{GeV}/m_H)^4 $. Consequently, the dominant SI scattering mechanisms are primarily driven by t-channel exchanges of both the SM-like Higgs $ h $ and the singlet Higgs $ h_s $. The $ h_s $-mediated process becomes particularly significant when a substantial mass hierarchy exists ($ m_{h_s} \ll m_h $), which is a key feature of this study. For sufficiently large values of \( m_H \), our analysis focuses on these two contributions, leading to the reformulation of Eqs.~(\ref{eqn0:BD}) and ~(\ref{eqn0:BU}):
\begin{align}
&{\cal{B}}_d\approx \left( \frac{125 GeV}{ m_{_{h}}} \right)^2 C_{ \tilde{\chi}_1^0 \tilde{\chi}_1^0 h}\,V_{h}^{\rm SM} -\left( \frac{125 GeV}{ m_{_{h_s}}} \right)^2 C_{ \tilde{\chi}_1^0 \tilde{\chi}_1^0 h_s}\,V_{h_s}^{\rm NSM} \tan\beta, \label{eqn1:BD}\\
&{\cal{B}}_u\approx \left( \frac{125 GeV}{ m_{_{h}}} \right)^2 C_{ \tilde{\chi}_1^0 \tilde{\chi}_1^0 h}\,V_{h}^{\rm SM},\label{eqn1:BU}
\end{align}
where we implement the approximations of $V_h^{\rm SM}$, $V_h^{\rm NSM}$, $V_{h_s}^{\rm SM}$, and $V_{h_s}^{\rm NSM}$ in the Higgs sector as described above, and consider the effect of a sufficiently large $\tan\beta$~\cite{Badziak:2015exr}.
Under these approximations, when wino decouples by eliminating the wino-related term proportional to $g_2 N_{12}$,  the generic CP-even scalar-neutralino-neutralino couplings in Eq.~(\ref{eq0:hichi01chi01}) reduce to the simplified expression:
\begin{align}
&C_{\tilde {\chi}^0_1 \tilde {\chi}^0_1 h }\sim \bigg[ \sqrt{2} \lambda N_{15} \left( N_{13} s_\beta + N_{14} c_\beta \right)
+ g_1 N_{11} \left( N_{13} c_\beta - N_{14} s_\beta \right) \bigg],
\label{eq1:hchi01chi01}
\\
&C_{\tilde {\chi}^0_1 \tilde {\chi}^0_1 h_s }\sim \bigg[\sqrt{2} \left( \lambda N_{13} N_{14} - \kappa N_{15} N_{15} \right)\bigg].
\label{eq1:hschi01chi01}
\end{align}
Substituting $N_{11}$, $N_{13}$, $N_{14}$ and $N_{15}$ of Eq.~(\ref{NIJ}) into Eqs.~(\ref{eq1:hchi01chi01}) and ~(\ref{eq1:hschi01chi01}), one finds
\begin{align}
&C_{\tilde {\chi}^0_1 \tilde {\chi}^0_1 h }\sim -\frac{\sqrt{2}m_Z^2s_W^4\mu_{eff}^5(M_2-m_{\tilde{\chi}^0_1})^2 }{vC_1}
\nonumber \\
&\Bigg[ \big(\frac{m_{\tilde{S}}}{\mu_{eff}} -\frac{m_{\tilde{\chi}^0_1}}{\mu_{eff}}\big)^2\big(\frac{m_{\tilde{\chi}^0_1}^2}{\mu_{eff}^2}-
1\big)\big(\frac{m_{\tilde{\chi}^0_1}}{\mu_{eff}}+s_{2\beta}\big)
\nonumber \\
&+2\bigg(\frac{\lambda v}{\mu_{eff}}\bigg)^2\big(\frac{m_{\tilde{S}}}{\mu_{eff}} -\frac{m_{\tilde{\chi}^0_1}}{\mu_{eff}}\big)\big(\frac{m_{\tilde{\chi}^0_1}^2}{\mu_{eff}^2}-
1\big)
+\bigg(\frac{\lambda v}{\mu_{eff}}\bigg)^4\big(\frac{m_{\tilde{\chi}^0_1}}{\mu_{eff}}-s_{2\beta}\big)\Bigg].
\label{eq2:hchi01chi01}
\\
\label{eq2:hschi01chi01}
&C_{\tilde {\chi}^0_1 \tilde {\chi}^0_1 h_s }\sim \frac{\sqrt{2}m_Z^2s_W^4\mu_{eff}^5(M_2-m_{\tilde{\chi}^0_1})^2}{v C_1}
\nonumber \\
& \Bigg[ \frac{\lambda v}{\mu_{eff}}\big(\frac{m_{\tilde{S}}}{\mu_{eff}} -\frac{m_{\tilde{\chi}^0_1}}{\mu_{eff}}\big)^2
 \Big[\big(\frac{m_{\tilde{\chi}^0_1}^2}{\mu_{eff}^2}+1\big)s_{\beta}c_{\beta}+\frac{m_{\tilde{\chi}^0_1}}{\mu_{eff}}\Big] +\frac{\kappa}{\lambda}\bigg(\frac{\lambda v}{\mu_{eff}}\bigg)^3c_{2\beta}^2
\nonumber \\
&+\bigg(\frac{\lambda v}{\mu_{eff}}\bigg)^3\big(\frac{m_{\tilde{S}}}{\mu_{eff}} -\frac{m_{\tilde{\chi}^0_1}}{\mu_{eff}}\big)\big(\frac{m_{\tilde{\chi}^0_1}}{\mu_{eff}}s_{2\beta}+1 \big)+\bigg(\frac{\lambda v}{\mu_{eff}}\bigg)^5s_\beta c_\beta \Bigg]
\end{align}
Substituting Eqs.~(\ref{eqn1:BD}),~(\ref{eqn1:BU}), ~(\ref{eq2:hchi01chi01}) and~(\ref{eq2:hschi01chi01}) into eq.~(\ref{eqn:FN}), one finds
\begin{align}\label{eqn2:FN}
f^{(N)}&\approx-\frac{F^{N}\sqrt{2}m_Z^2s_W^4(M_2-m_{\tilde{\chi}^0_1})^2 }{vC_1} \mu_{eff}^5 \nonumber \\
&\Bigg\{ \left( \frac{125 GeV}{ m_{_{h}}} \right)^2 2\,V_{h}^{\rm SM} \Bigg[ \bigg(\frac{m_{\tilde{S}}}{\mu_{eff}} -\frac{m_{\tilde{\chi}^0_1}}{\mu_{eff}}\bigg)^2\bigg(\frac{m_{\tilde{\chi}^0_1}^2}{\mu_{eff}^2}-
1\bigg)\bigg(\frac{m_{\tilde{\chi}^0_1}}{\mu_{eff}}+s_{2\beta}\bigg)
\nonumber \\
&+2\bigg(\frac{\lambda v}{\mu_{eff}}\bigg)^2\bigg(\frac{m_{\tilde{S}}}{\mu_{eff}} -\frac{m_{\tilde{\chi}^0_1}}{\mu_{eff}}\bigg)\bigg(\frac{m_{\tilde{\chi}^0_1}^2}{\mu_{eff}^2}-
1\bigg)+\bigg(\frac{\lambda v}{\mu_{eff}}\bigg)^4 \bigg(\frac{m_{\tilde{\chi}^0_1}}{\mu_{eff}}-s_{2\beta}\bigg)\Bigg] \nonumber \\
&+ \left( \frac{125 GeV}{ m_{_{h_s}}} \right)^2 \,V_{h_s}^{\rm NSM} \tan\beta
 \Bigg[ \frac{\lambda v}{\mu_{eff}} \bigg(\frac{m_{\tilde{S}}}{\mu_{eff}} -\frac{m_{\tilde{\chi}^0_1}}{\mu_{eff}}\bigg)^2
 \Big[\bigg(\frac{m_{\tilde{\chi}^0_1}^2}{\mu_{eff}^2}+1 \bigg) s_{\beta}c_{\beta}+\frac{m_{\tilde\chi^0_1}}{\mu_{eff}}\Big]
\nonumber \\
&+\bigg(\frac{\lambda v}{\mu_{eff}}\bigg)^2\kappa v c_{2\beta}^2+\bigg(\frac{\lambda v}{\mu_{eff}}\bigg)^3\bigg(\frac{m_{\tilde{S}}}{\mu_{eff}} -\frac{m_{\tilde{\chi}^0_1}}{\mu_{eff}}\bigg)\bigg(\frac{m_{\tilde{\chi}^0_1}}{\mu_{eff}}s_{2\beta}+1 \bigg)+\bigg(\frac{\lambda v}{\mu_{eff}}\bigg)^5s_\beta c_\beta \Bigg]
 \Bigg\}
\end{align}
where the assumption $F^{(N)}_d \approx F^{(N)}_u \approx F^{(N)}$ is applied. Given $m_{\tilde{\chi}^0_1} \neq \, M_2, \mu_{eff}, m_{\tilde{S}}$, and $C_1 \neq 0$ for a bino-dominated LSP, a blind spot for the SI cross-section arises when the quantity within the curly brackets on the right-hand side of Eq.~(\ref{eqn2:FN}) vanishes. This condition signifies the possibility of significant cancellation between the $h$- and $h_s$-mediated contributions. Such cancellation mechanisms become effective only under a substantial mass hierarchy between $m_{h_s}$ and $m_h$ (i.e. when $m_{h_s} \ll m_h$). Furthermore, when a blind spot for the \( h \) contribution in Eq.~(\ref{eqn2:FN}) occurs, the \( h_s \) contribution becomes significant regardless of other factors.

%\subsection{Dark matter relic density}
The measured DM relic density is $\Omega_{\rm DM} h^2 \simeq 0.12$~\cite{Planck:2018vyg}, where $h \equiv H_0/(100\,{\rm km/s/Mpc})$ denotes the normalized Hubble constant, and $\Omega_{\rm DM} \equiv \rho_{\rm DM}/\rho_c$ is the DM density in units of the critical density. As a viable DM candidate, the relic density of $\tilde{\chi}_1^0$, which we denote as $\Omega_{\tilde{\chi}_1^0} \equiv \rho_{\tilde{\chi}_1^0}/\rho_c$, should match the observed value $\Omega_{\rm DM}$. For DM produced via standard thermal freeze-out~(FO), the relic density and the effective (thermally averaged) annihilation cross section at freeze-out, $\left\langle \sigma v \right\rangle_{\rm FO}$, are approximately related as
\begin{equation} \label{eq:relic_density}
 \Omega_{\tilde{\chi}_1^0} h^2 \sim 0.1 \times \frac{3 \times 10^{-26}\,{\rm cm}^3/{\rm s}}{\left\langle \sigma v \right\rangle_{\rm FO}} \;.
\end{equation}
Since at the time of freeze-out (parameterized by the temperature $x_F \equiv m_{\tilde{\chi}_1^0}/T$) the DM candidate is usually non-relativistic (typically, $x^{\rm FO} \sim 20$), one often expands $\left\langle \sigma v \right\rangle_{\rm FO}$ as
\begin{equation} \label{eq:annxsec_NR}
 \left\langle \sigma v \right\rangle_{x_F}= a + 6 b / {x_F} + \mathcal{O}({x_F}^{-2})\;.
\end{equation}
In scenarios featuring relatively light singlet-like scalar or pseudoscalar states, a bino-dominated DM candidate can attain the observed relic density either through resonant annihilation via scalar funnels or through co-annihilation with a singlino-dominated neutralino~\cite{Abdallah:2020yag}. If the LSP possesses a sizable higgsino admixture, efficient annihilation channels into electroweak gauge bosons and doublet-like Higgs bosons become accessible via $s$-channel $Z$ or Higgs exchange, as well as through $t$-channel neutralino and chargino exchange. For heavier LSPs ($m_{\chi_1^0} > m_t$), annihilation into top-quark pairs through Higgs exchange can  provide a significant contribution. In the large $\tan \beta$ regime,
where Higgs couplings to $b \bar b$ and $\tau^+ \tau^-$ are significantly enhanced, $b \bar b$ and $\tau^+ \tau^-$ final states are possible via
$s$-channel Higgs exchange (not necessarily resonant). Additionally, when extra scalar or pseudoscalar particles are sufficiently light, further annihilation channels---$Z\,h_s$, $h_s\,h_s$, $h_s\,a_s$, and $a_s\,a_s$---may become kinematically accessible. These processes, mediated through $s$-channel exchange of $Z$, $h_s$, $a_s$, or via $t$-channel exchange of heavier neutralinos, can play a critical role in reducing the neutralino relic density. The corresponding annihilation rates are sensitive to the NMSSM-specific couplings $\lambda$ and $\kappa$, which govern the interactions of the singlet sector.

\section{\label{sec:NR}Numerical Results}

\subsection{\label{sec:scan} Research strategy}
We used the \texttt{NMSSMTools-6.0.0} package~\cite{Ellwanger:2004xm,Ellwanger:2005dv} to perform a comprehensive scan over the following parameter space:
\begin{eqnarray}\label{Parameters_scan}
&& 0.0 < \lambda\leq 0.75,\quad |\kappa| \leq 0.75, \quad 1\leq\tan\beta\leq 60,
\nonumber\\
&& \quad 0.1{\rm ~TeV}< |\mu| \leq 1.5 {\rm ~TeV},\quad |A_{\lambda}| \leq 5 {\rm ~TeV},\quad |A_{\kappa}| \leq 1.5 \, {\rm TeV},\nonumber\\
&& | M_1 |\leq 1.5 {\rm ~TeV}, \quad 0.1{\rm ~TeV}\leq M_2 \leq 1.5 {\rm ~TeV}, \nonumber\\
&& |A_{t}|= |A_{b}|\leq 5 {\rm ~TeV}, \quad 0.1{\rm ~TeV}\leq M_{\tilde{l}_L}, M_{\tilde{l}_R}\leq 1.5 {\rm ~TeV}, \nonumber
\end{eqnarray}
where all parameters were defined at the scale $Q = 1~{\rm TeV}$. To obtain an SM-like Higgs boson mass ($m_h\approx125 ~\rm GeV$), the soft trilinear coefficients $A_t$ and $A_b$ were treated as free and equal to adjust the Higgs mass spectrum. The masses of the left (right) three generation sleptons  take
a common value $ M_{\tilde{l}_L}(M_{\tilde{l}_R}$), which served as free parameters to help explain the muon $g-2$ anomaly, and to relax LHC constraints. Other less relevant parameters were fixed at 3 TeV, including the gluino mass $M_3$, all three generations of squark masses, and the soft trilinear couplings apart from $A_t$ and $A_b$. We also imposed the condition $N_{11}^2 > 0.5$ to ensure a bino-dominated lightest neutralino $\tilde\chi^0_1$.

In order to ensure the conclusions drawn in this work are as comprehensive and robust as possible, we employ the MultiNest algorithm presented in \cite{Feroz:2008xx,Feroz:2013hea} during the parameter scan, with ${\it nlive}$ = 5000\footnote{The $n_{\rm live}$ parameter defines the number of live points sampled  for determining the iso-likelihood contour in each iteration of the scan.}. The result of the scan  includes the Bayesian evidence, which is defined as
\begin{eqnarray}
Z(D|M) \equiv \int{P(D|O(M,\Theta)) P(\Theta|M) \prod d \Theta_i}. \nonumber
\end{eqnarray}
Here, $P(\Theta|M)$ represents the prior probability density function (PDF) for the parameters $\Theta = (\Theta_1,\Theta_2,\cdots)$ within the model $M$, and $P(D|O(M, \Theta)) \equiv \mathcal{L}(\Theta)$ denotes the likelihood function for the theoretical predictions on observables $O$ compared to their experimentally measured values $D$. From a computational standpoint, the Bayesian evidence acts as an average likelihood, with its value determined by the priors of the model parameters. For various scenarios under a given theoretical framework, a larger $Z$ indicates better agreement between the scenario and the data. The output further includes weighted and unweighted parameter samples, whose distribution are governed by the posterior PDF $P(\Theta|M,D)$, expressed as:
	\begin{equation}\label{eq:bayes}
	\begin{aligned}
	P(\Theta|M,D)=\frac{P(D|O(M,\Theta))P(\Theta|M)}{Z}.
	\end{aligned}
	\end{equation}
This equation represents our current understanding of the parameters $\Theta$ based on the experimental data $D$, or equivalently, the updated prior PDF following the incorporation of the experimental data. While this quantity may exhibit sensitivity to the prior's shape, such sensitivity can be mitigated by adequate data, rendering it negligible under specific circumstances~\cite{Bayes}. Consequently, one can deduce the inherent physics of the model through the distribution of the samples.
The likelihood function constructed in this study comprises:
\begin{eqnarray}
\mathcal{L}=\mathcal{L}_{\rm{m_{h}}} \times \mathcal{L}_{h, {\rm coupling}} \times {\mathcal{L}_{B} } \times {\mathcal{L}_{\Omega h^2} }\times {\mathcal{L}_{DM}} \times \mathcal{L}_{\Delta a_\mu},
\label{Likelihood}
\end{eqnarray}
where
\begin{itemize}
\item $\mathcal{L}_{\rm{m_{h}}}$ and $\mathcal{L}_{h, {\rm coupling}}$ correspond to the likelihoods for the measured Higgs boson mass and couplings, respectively. The Higgs mass $m_h$ was computed including leading electroweak corrections, two-loop effects, and propagator corrections following Ref. \cite{Degrassi:2009yq}, with a central experimental value $m_h = 125.09{\rm GeV}$\cite{Aad:2015zhl}, and an assumed total (experimental + theoretical) uncertainty of $3{\rm GeV}$. $\mathcal{L}_{h, {\rm coupling}}$ was constructed using a seven-parameter $\kappa$-framework, incorporating experimental central values, uncertainties, and correlations from ATLAS Run-II results (80fb$^{-1}$)\cite{Aad:2019mbh}, with statistical methods informed by Ref. \cite{ParticleDataGroup:2024cfk}.

\item ${\mathcal{L}_{B} }$ represents the likelihood function for the measured  branching ratio of the $B \to X_s \gamma$ and $B_s \to \mu^+\mu^-$. These ratios were calculated by the methods from  in Refs.~\cite{Domingo:2007dx,Domingo:2015wyn} and were constrained within the $2\sigma$ experimental bounds~\cite{ParticleDataGroup:2024cfk}.

\item ${\mathcal{L}_{\Omega h^2}}$ and ${\mathcal{L}_{DM}}$ describe constraints from the DM relic density and direct detection cross-sections. The relic density likelihood was based on the Planck 2018 result, $\Omega h^2 = 0.120$\cite{Planck:2018vyg}, with a $20\%$ theoretical uncertainty ($0.096 \leq \Omega h^2 \leq 0.144$). For direct detection, a Gaussian likelihood centered at zero was used, with uncertainty defined as $\delta_{\sigma}^2 = {\rm UL}\sigma^2 / 1.64^2 + (0.2 \sigma)^2$, where ${\rm UL}\sigma$ denotes the 90\% C.L. upper limits from the LZ-2022 experiment\cite{LZ:2022lsv}, and $0.2\sigma$ accounts for theoretical uncertainties~\cite{Matsumoto:2016hbs}. These observables were computed using the \texttt{micrOMEGAs} package~\cite{Belanger:2004yn,Belanger:2006is,Belanger:2013oya,Belanger:2018ccd}.
\item $\mathcal{L}_{\Delta a_\mu}$ accounts for the likelihood function of the muon $g-2$ anomaly, and is given by
\begin{eqnarray}
\mathcal{L}_{\Delta a_\mu} &=& \exp \left[ -\frac{\chi^2_{\Delta a_\mu}}{2} \right] = \exp \left [ -\frac{1}{2} \left( \frac{a^{\rm SUSY}_\mu - 2.49\times 10^{-9}}{4.8\times 10^{-10}}\right)^2 \right]
\label{chi2-excesses}
\end{eqnarray}
where $\Delta a_\mu \equiv a_\mu^{\rm Exp} - a_\mu^{\rm SM}$ denotes the deviation between experimental central value of $a_\mu$ and its SM prediction. Incorporating the most recent measurement by the Fermilab Muon $g-2$ experiment~\cite{Muong-2:2023cdq}, the combined experimental average $a_\mu^{\rm Exp}$ exhibits a $5.1\sigma$ discrepancy from the SM prediction $a_\mu^{\rm SM}$ as reported by the Muon $g-2$ Theory Initiative~\cite{Aoyama:2020ynm},
\begin{eqnarray}
\Delta a_{\mu}=a_{\mu}^{\rm Exp}-a_{\mu}^{\rm SM}=\left(2.49\pm 0.48\right)\times 10^{-9}.
\label{eq:muong-2}
\end{eqnarray}
which implies the presence of potential new physics effects. While it is important to recognize that uncertainties in the calculation of leading-order hadronic contributions may account for part of the observed deviation~\cite{Stoffer:2023gba,Bray-Ali:2023whf}~\footnote{ The potential underestimation of theoretical uncertainties remains a topic of ongoing debate, thereby complicating a definitive comparison between the experimental measurement of the muon $g-2$ and theoretical predictions~\cite{Venanzoni:2023mbe, Kuberski:2023qgx, Muong-2:2024hpx}. Notably, the Budapest-Marseille-Wuppertal collaboration performed lattice calculations of the leading-order hadronic vacuum polarization contribution with sub-percent precision~\cite{Borsanyi:2020mff}, reducing the observed discrepancy to approximately 1.5$\sigma$ ~\cite{Davier:2023cyp}. Furthermore, a dispersive analysis based on the latest $e^+e^- \to \pi^+\pi^-$ cross section measurements by the CMD-3 experiment demonstrates good agreement with the combined FermiLab-BNL result, although it remains in tension with all previous measurements~\cite{CMD-3:2023alj,CMD-3:2023rfe}.}, significant attention has been directed toward investigating potential BSM explanations for this anomaly~\cite{Athron:2021iuf,Lindner:2016bgg}.
\end{itemize}

The acquired samples were further refined according to the following criteria: First, the SM-like Higgs boson mass was required to lie within the range of 122--128 GeV. Additionally, the p-value obtained using the ~\textsf{HiggsSignals-2.6.2} code~\cite{Bechtle:2013xfa,Stal:2013hwa,Bechtle:2014ewa,Bechtle:2020uwn} had to be greater than 0.05, ensuring compatibility of the SM-like Higgs properties with measurements reported by the ATLAS and CMS collaborations at the $95\%$ confidence level. Constraints from searches for additional Higgs bosons were also imposed using \textsf{HiggsBounds-5.10.2}~\cite{Bechtle:2008jh,Bechtle:2011sb,Bechtle:2012lvg,Bechtle:2013wla,Bechtle:2020pkv}. Furthermore, the condition $\kappa^2+\lambda^2<0.5$ was applied to ensure perturbativity of the theory up to the scale of $10^{16}$ GeV~\cite{Miller:2003ay}. Finally, the DM relic abundance was required to fall within $\pm 10\%$ of the Planck 2018 central value $\Omega h^2 = 0.12$ ( i.e., $0.108 \leq \Omega h^2 \leq 0.132$)~\cite{Planck:2018vyg}.

%%%%%%%%%%%%%%%%%%%%%%%%%% F I G U R E %%%%%%%%%%%%%%%%%%%%%%%%%%%%%%%%%%%%%%%
\begin{figure}[h]
\begin{center}
\includegraphics[width=0.9\textwidth]{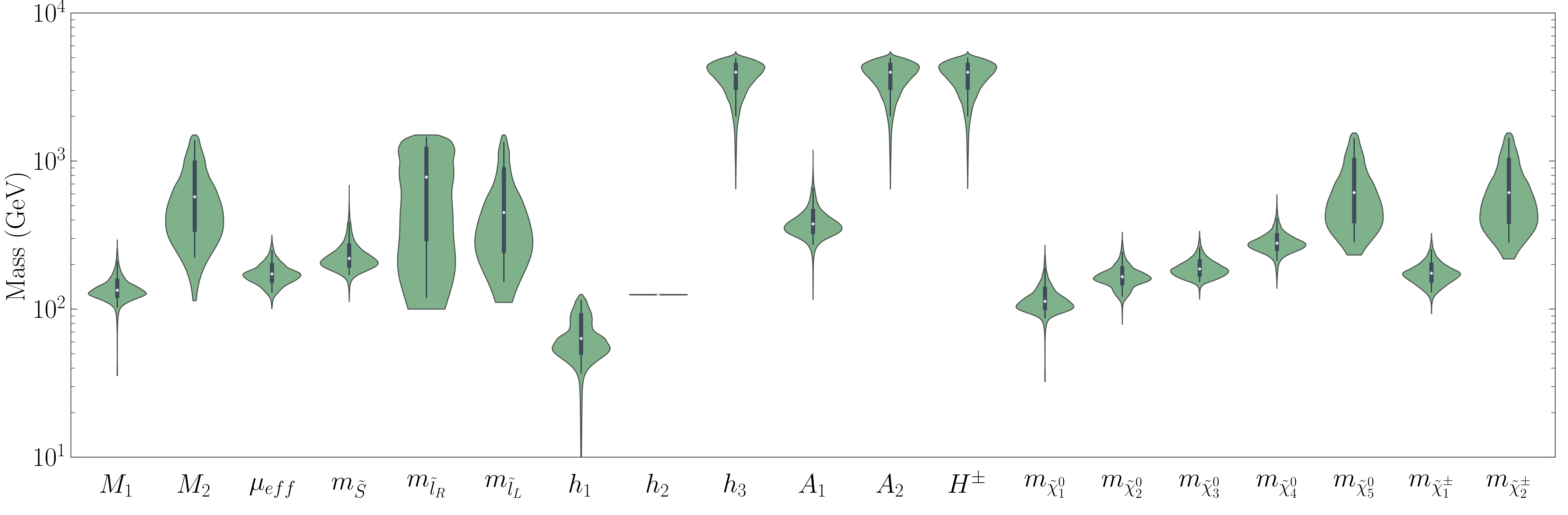}
\vspace{-0.4cm}
\caption{\label{fig:masses}Violin plots showing mass distributions of Higgs bosons, EWinos, and sleptons for the refined samples. The violins are scaled by count. The thick vertical bar at the center indicates the interquartile range, with the white dot representing the median; the long vertical line shows the $95\%$ confidence interval.
}
\end{center}
\vspace{-1em}
\end{figure}
%%%%%%%%%%%%%%%%%%%%%%%%%% F I G U R E %%%%%%%%%%%%%%%%%%%%%%%%%%%%%%%%%%%%%%%

%%%%%%%%%%%%%%%%%%%%%%%%%% F I G U R E %%%%%%%%%%%%%%%%%%%%%%%%%%%%%%%%%%%%%%%
\begin{figure}[h]
\begin{center}
\resizebox{1.\textwidth}{!}{
\includegraphics[width=0.5\textwidth]{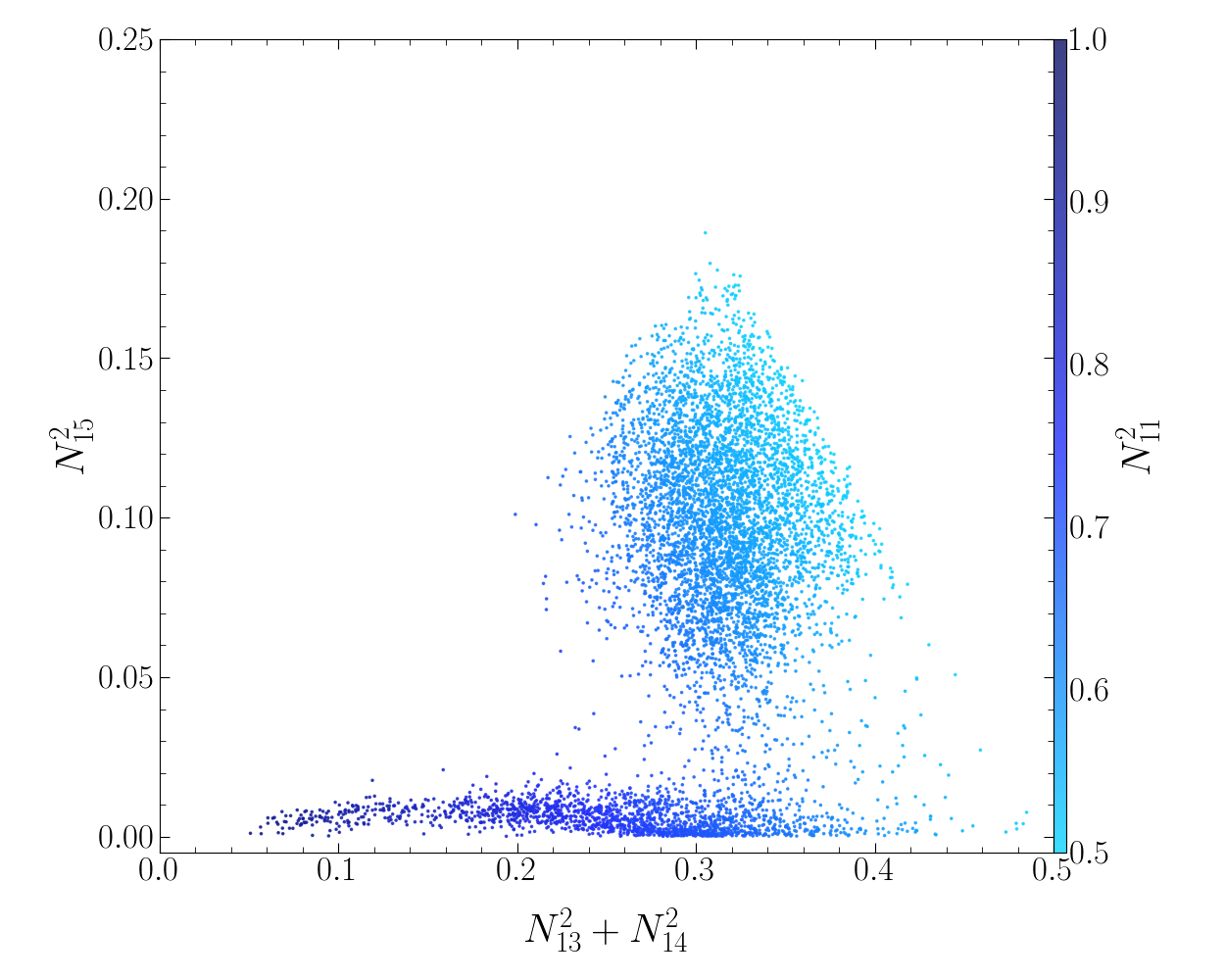}\hspace{-0.4cm}
\includegraphics[width=0.5\textwidth]{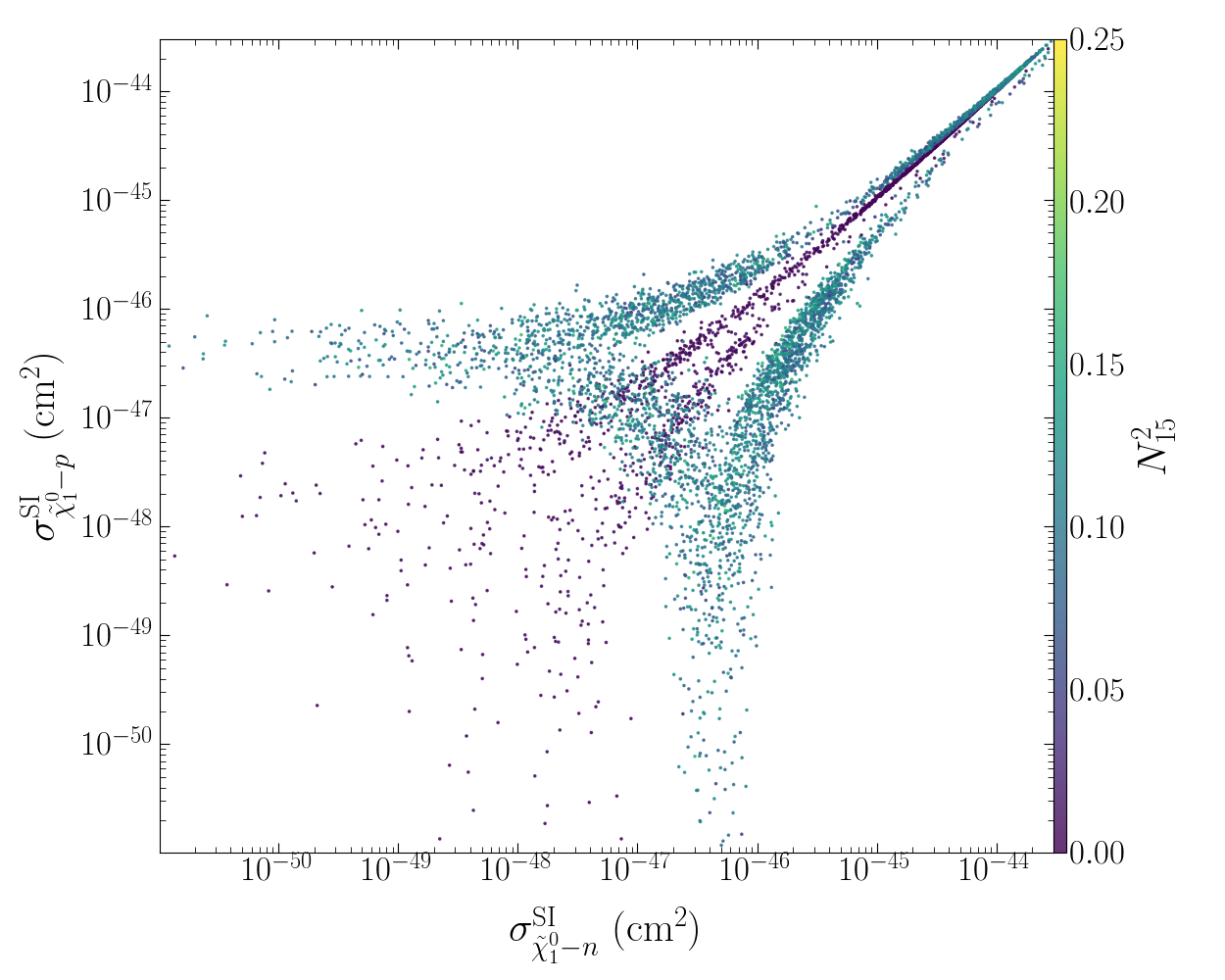}
}
\vspace{-0.9cm}
\caption{\label{fig2}Left: Components of ${\tilde{\chi}_1^0}$ for the refined samples. Right: Projection of the refined samples onto the $\sigma_{\tilde{\chi}_1^0-p}^{SI}-\sigma_{\tilde{\chi}_1^0-n}^{SI}$ plane, with colors representing the values of $N_{15}^2$.
}
\end{center}
\end{figure}
%%%%%%%%%%%%%%%%%%%%%%%%%% F I G U R E %%%%%%%%%%%%%%%%%%%%%%%%%%%%%%%%%%%%%%%

%%%%%%%%%%%%%%%%%%%%%%%%%%%%%%%%%%%%%%%%%%%%%%%%%%%%%%%%%%%%%%%%%%%%%%%%%%%%%%%%%%%%%%%%%%%%%%%
\begin{figure*}[h!]
		\centering
		\resizebox{1.\textwidth}{!}{
 \includegraphics[width=0.90\textwidth]{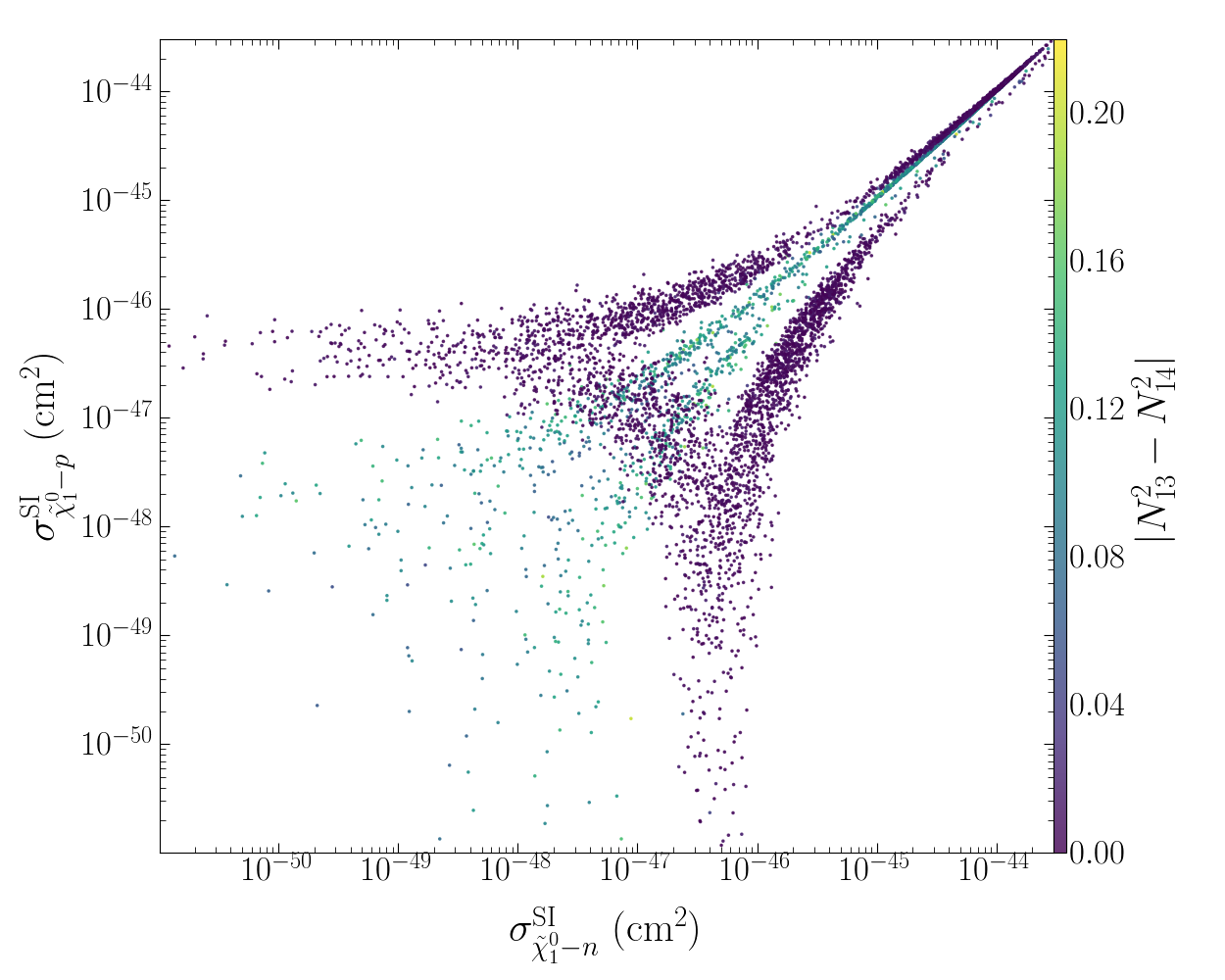}\hspace{-0.4cm}
 \includegraphics[width=0.90\textwidth]{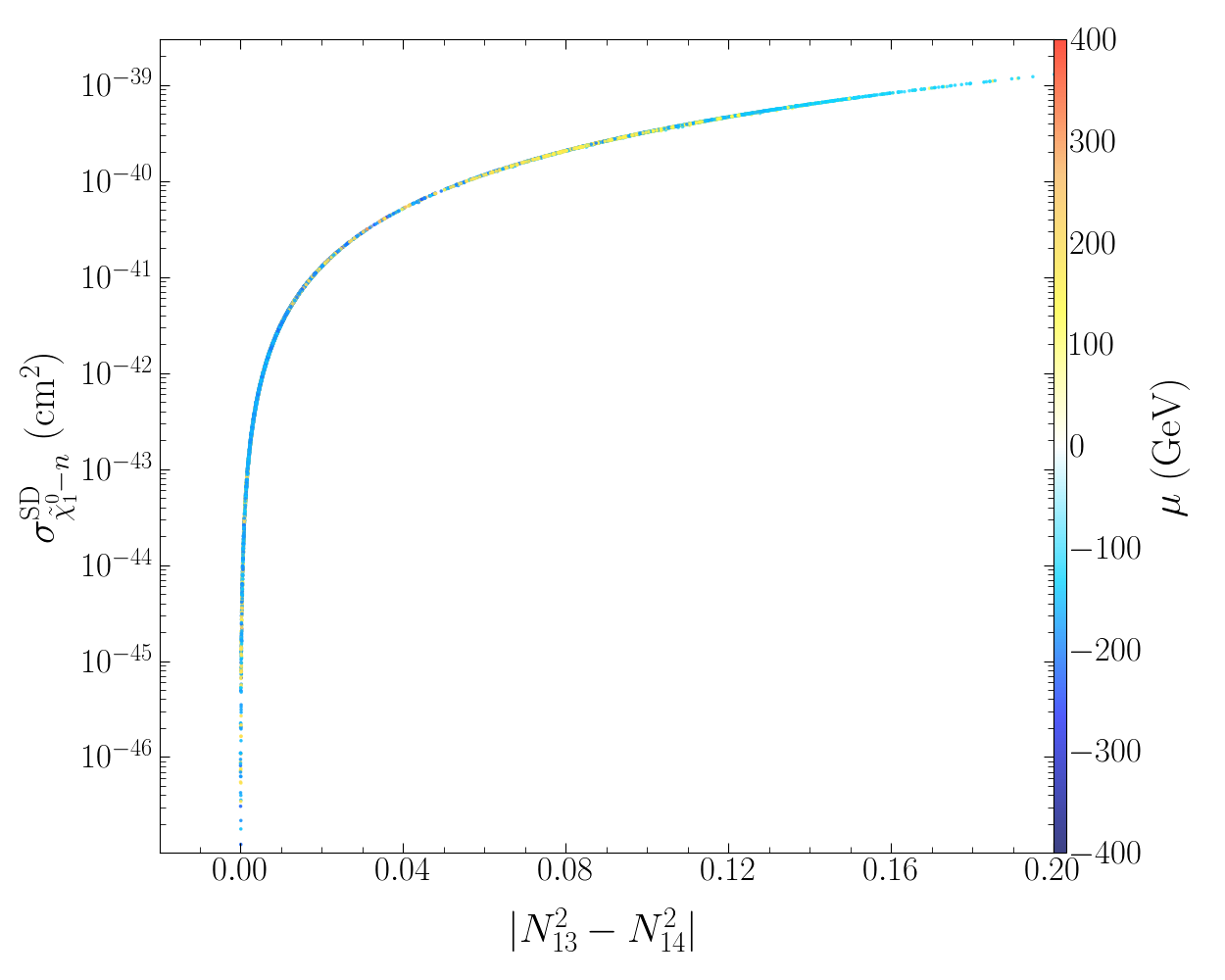}
 }
 \vspace{-0.8cm}
 \caption{Left: Projection of the refined samples onto the $\sigma_{\tilde{\chi}_1^0-p}^{SI}-\sigma_{\tilde{\chi}_1^0-n}^{SI}$ plane, with colors representing the values $|N_{13}^2-N_{14}^2|$. Right: Projection onto the $\sigma_{\tilde{\chi}_1^0-n}^{SD}- |N_{13}^2-N_{14}^2|$ plane, with colors denoting values of $\mu_{eff}$. \label{fig3} }
\end{figure*}	
%%%%%%%%%%%%%%%%%%%%%%%%%%%%%%%%%%%%%%%%%%%%%%%%%%%%%%%%%%%%%%%%%%%%%%%%%%%%%%%%%%%%%%%%%%%%%%%	

For the refined samples, the mass distributions of the Higgs and SUSY particles are shown in Fig.~\ref{fig:masses} using violin plots \footnote{A violin plot combines the features of a box plot and a probability density distribution plot~\cite{hintze1998violin}.}. From Fig.~\ref{fig:masses}, it can be observed that ${h_2}$ corresponds to the SM-like Higgs boson ($h$), and $m_{h_1}\leq m_{h_2} \ll m_{h_3}$, indicating the presence of a light singlet-like scalar state ($h_s \equiv h_1$). In this scenario, as discussed in Section \ref{DM} on DM-nucleon cross sections, the contribution from the non-SM doublet Higgs boson ($H \equiv h_3$) to the SI cross section is suppressed by a factor of $(125/m_{H})^4$. Furthermore, the contribution to the SI cross section from the $h_s$-mediated process may become particularly significant due to the large mass hierarchy between $m_{h_s}$ and $m_{h}$.
Regarding the masses of the EWinos, the parameter ranges are as follows: $M_{1} \in (-291,-35){~\rm GeV}~ \bigcup~ (36,293){~\rm GeV}$, $\mu_{eff} \in (-311,-101){~\rm GeV}~ \bigcup~ (102,305){~\rm GeV}$, $m_{\tilde{S}} \in (-687,-125){~\rm GeV}~ \bigcup~ (112,564){~\rm GeV}$, $M_{2} \in (114,1500){~\rm GeV}$, $|\mu_{eff}/M_1| \in (1.2,4.53) $, $|m_{\tilde{S}}/\mu_{eff}| \in (1.0,3.9) $, $|m_{\tilde{\chi}_1^0}/M_1| \in (0.71,0.98) $. $|m_{\tilde{\chi}_2^0}/\mu_{eff}| \in (0.7,1.13) $, $|m_{\tilde{\chi}_3^0}/\mu_{eff}| \in (1.0,1.33) $, $|m_{\tilde{\chi}_4^0}/m_{\tilde{S}}| \in (0.7,1.5) $, and $ |m_{\tilde{\chi}_1^{\pm}}/\mu_{eff}|\in (0.7,1.0)$. These results indicate the presence of one light chargino and four relatively light neutralinos, with the lightest of them (the LSP) being bino-dominant.

The composition of ${\tilde{\chi}_1^0}$ is illustrated in the left panel of Fig.~\ref{fig2}. The data points can be categorized into two systems based on the singlino component of ${\tilde{\chi}_1^0}$: a bino-higgsino system ($N_{15}^2 < 0.05$) and a bino-higgsino-singlino system ($N_{15}^2 > 0.05$). From the right panel of Fig. \ref{fig2} and Fig. \ref{fig3}, it is evident that the DM-nucleon cross sections exhibit significant differences between the two DM systems.

In the next section, we will explore the specific challenges related to the direct detection, indirect detection, and relic density constraints of DM for the refined samples.

%%%%%%%%%%%%%%%%%%%%%%%%%% F I G U R E %%%%%%%%%%%%%%%%%%%%%%%%%%%%%%%%%%%%%%%
\begin{figure}[t]
\begin{center}
\resizebox{1.\textwidth}{!}{
\includegraphics[width=0.51\textwidth]{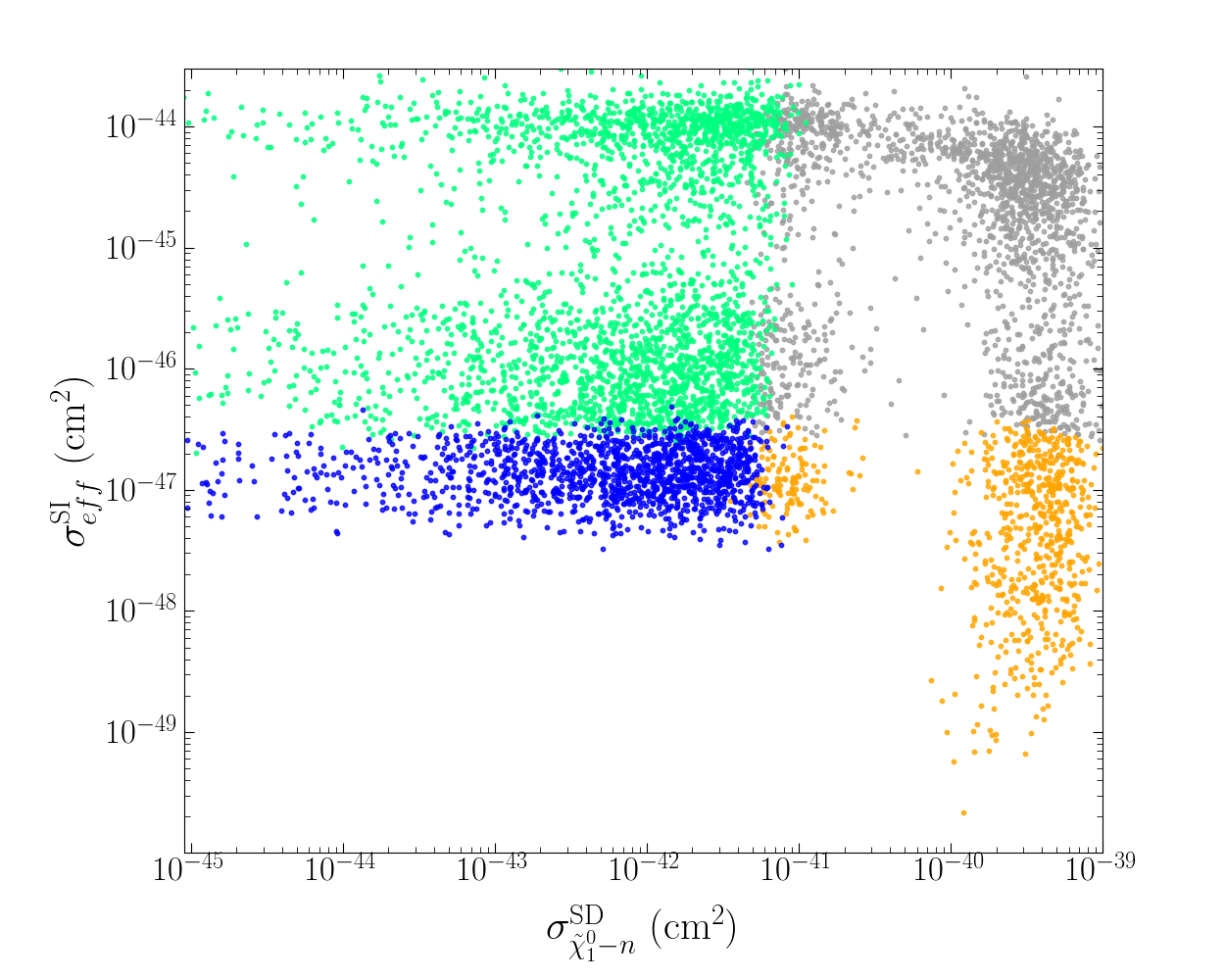}\hspace{-0.4cm}
\includegraphics[width=0.5\textwidth]{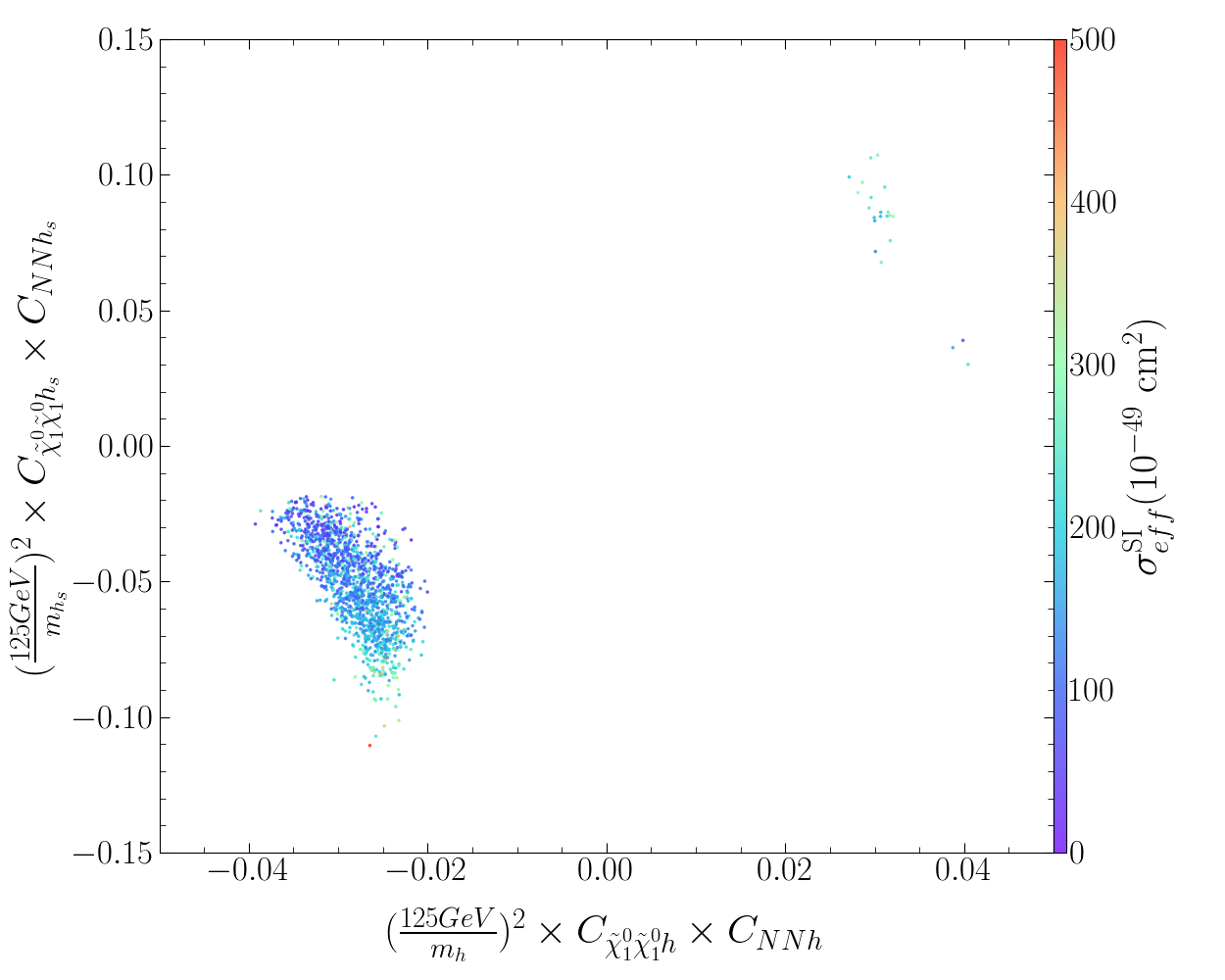}
}
\vspace{-0.8cm}
\caption{\label{fig4}In the left panel, the refined samples are projected onto the $\sigma_{eff}^{SI}-\sigma_{\tilde{\chi}_1^0-n}^{SD}$ plane. Grey samples are excluded by both the LZ-2022 limits on SI and SD cross sections. Orange and green samples are excluded by either the SI or SD constraints, respectively, while blue samples remain viable under all constraints. In the right panel, the surviving blue samples from the left panel are projected onto the $(\frac{125 \text{ GeV}}{m_{h_s}})^2 \times C_{{\tilde{\chi}^0_{1}}{\tilde{\chi}^0_{1}}h_s} \times C_{N N h_s} - (\frac{125 \text{ GeV}}{m_{h}})^2 \times C_{{\tilde{\chi}^0_{1}}{\tilde{\chi}^0_{1}} h} \times C_{N N h}$ plane. The color bar indicates the values of $\sigma_{eff}^{SI}$.
}
\end{center}
\end{figure}
\subsection{\label{DM_result} Dark Matter phenomenology}

 As illustrated in Fig.~\ref{fig3}, the bino-higgsino system is significantly constrained by the LZ experiment through the SD cross-section limits. This is because $\sigma^{SD}$ is proportional to $|N_{13}^2-N_{14}^2|$, and the considered samples exhibit large values of this quantity. Additionally, we examined the composition of other neutralinos and found that ${\tilde{\chi}_{2,3}^0}$ and $\tilde{\chi}_1^{\pm}$ are higgsino-dominated, and ${\tilde{\chi}_4^0}$ is singlino-dominated. Consequently, the DM phenomenology at the LHC largely resembles that of the MSSM with a higgsino-like NLSP ~\cite{Dutta:2014hma}. In this context, $\sigma^{SD}$ is suppressed by $1/\mu_{eff}^4$, and the LZ-2022 experiment places a lower bound of $\mu_{eff} > 380$ GeV \cite{He:2023lgi}.
In contrast, within the bino-higgsino-singlino system, the values of $|N_{13}^2-N_{14}^2|$ approach zero, satisfying the blind-spot condition described in Eq.(~\ref{eq1:zchi10chi10}). This results in extremely small SD cross-sections, with $\sigma_{\tilde{\chi}_1^0-n}^{SD}$ reaching as low as $10^{-50}$ cm$^2$. A moderate to large value of `$\lambda$' ($0.3 < \lambda < 0.7$) plays a critical role in facilitating sufficient mixing of a bino LSP with higgsino (and consequently singlino) components. We carefully analyzed this scenario and found that for $1<m_{\tilde{S}}/ m_{\tilde{\chi}_1^0} <3 $ where $m_{\tilde{S}}$ and $m_{\tilde{\chi}_1^0}$ share the same sign and the cancellation in Eq.(\ref{eq1:zchi10chi10}) occurs between the first term and the sum of the remaining two terms---the singlino mass is not much larger than the LSP mass. However, this induces a significant bino-singlino mixing in $\tilde{\chi}_1^0$, which may conflict with SI cross-section constraints, as shown in the $\sigma_{\tilde{\chi}_1^0-p}^{SI}$ vs. $\sigma_{\tilde{\chi}_1^0-n}^{SI}$ panels of Figs.~\ref{fig2} and~\ref{fig3}. This represents a distinct feature of the NMSSM, where the bino-dominated LSP is tempered by singlino mixing that necessarily involves the higgsinos.

It is worth noting that when considering the LZ-2022 limits on both SI and SD cross sections, the SI cross sections for proton and neutron at the survival points often show that when one value is significantly higher, its counterpart may be relatively lower. Thus, $\sigma_{eff}^{SI}$ in
Eq.(\ref{Effective-Cross-Section}) is calculated, and the refined samples are plotted onto the $\sigma_{eff}^{SI}-\sigma_{\tilde{\chi}_1^0-n}^{SD}$ panel of Fig.~\ref{fig4}. Grey points are excluded by both the LZ-2022 SI and SD limits. Orange and green samples are excluded by one of the two limits, while blue samples remain viable under all constraints. However, these blue samples are ruled out by the recent WIMP search results from LZ (2024)~\cite{LZ:2024zvo}, which appear to exclude thermally produced natural SUSY WIMPs.

For the remaining samples, the contributions to the SI cross-section from $t$-channel exchange of the SM-like Higgs boson $h$ and the singlet Higgs boson $h_s$ are shown in the right panel of Fig.\ref{fig4}, with the color bar indicating the values of $\sigma_{\text{eff}}^{\text{SI}}$. One can see that the $t$-channel exchange of the singlet Higgs boson $h_s$ is the dominant contribution. Large values of $|(\frac{125 \text{ GeV}}{m_{h_s}})^2 \times C_{{\tilde{\chi}^0_{1}}{\tilde{\chi}^0_{1}}h_s} \times C_{N N h_s}|$ can result in an enhanced $\sigma_{eff}^{SI}$. Table\ref{table-1} summarizes the parameter ranges of the samples that satisfy the LZ limits on the SI and SD cross sections. Since $h_s$ is significantly lighter than $h$, the primary contribution to the SI scattering cross section may originate from the $t$-channel exchange of $h_s$.

\begin{table}[h]
\caption{Parameter ranges for the samples that satisfy the LZ-2022 limits on both SI and SD cross sections. All quantities with mass dimension are in units of ${\rm GeV}$. \label{table-1}}

\centering

\vspace{0.2cm}

\resizebox{1\textwidth}{!}{%
\renewcommand{\arraystretch}{1.5}
\begin{tabular}{c|c|c|c|c|c}
\cline{1-6}
para & range & para & range & para & range \\
\cline{1-6} \multicolumn{1}{c|}
{\multirow{1}{*} {~$\tan \beta $~}} & $9\thicksim 34$ & \multicolumn{1}{c|}{\multirow{1}{*}{~$A_{t}$~}}
& $-2650 \thicksim 2250$ & \multicolumn{1}{c|}{\multirow{1}{*}{~$m_{\tilde{\chi}_1^0}$~}}& ($-212\thicksim-70$)$\cup$($93\thicksim 160$) \\
%%%\cline{2-6}
%\multicolumn{1}{|c|}{}& $7\thicksim 18$& \multicolumn{1}{|c|}{}
%& $65\thicksim85$& \multicolumn{1}{|c|}{} & $-400\thicksim -60$ \\
\cline{1-6} \multicolumn{1}{c|}{\multirow{1}{*}{~$\kappa$~}}
& ($-0.39\thicksim-0.18$)$\cup$($0.16\thicksim0.43$) & \multicolumn{1}{c|}{\multirow{1}{*}{~$m_{h_1}$~}}
& $33 \thicksim 75 $& \multicolumn{1}{c|}{\multirow{1}{*}{~$ m_{\tilde{\chi}_2^0}$~}} & ($-270\thicksim-130$)$\cup$($118\thicksim 221$) \\
%%%\cline{2-6}
%\multicolumn{1}{|c|}{}& $0.15\thicksim0.49$& \multicolumn{1}{|c|}{}& $125\thicksim195$
%& \multicolumn{1}{|c|}{}& $45\thicksim 120 $ \\
\cline{1-6}\multicolumn{1}{c|}{\multirow{1}{*}{~$\lambda$~}}
& $0.31\thicksim 0.63$& \multicolumn{1}{c|}{\multirow{1}{*}{~$m_{h_2}$~}}
& $125\thicksim126$ & \multicolumn{1}{c|}{\multirow{1}{*}{~$m_{\tilde \chi_3^0}$~}}
& ($-255\thicksim-146$)$\cup$($144\thicksim 298$) \\
%%%\cline{2-6}
%\multicolumn{1}{|c|}{}& $0.28\thicksim 0.68$& \multicolumn{1}{|c|}{} & $150 \thicksim260$
%& \multicolumn{1}{|c|}{}& $120\thicksim 350$ \\
\cline{1-6}\multicolumn{1}{c|}{\multirow{1}{*}{~$\mu$~}}& ($-283\thicksim-128$)$\cup$($144\thicksim 254$)
& \multicolumn{1}{c|}{\multirow{1}{*}{~$m_{h_3}$~}}
& $1640 \thicksim 5190 $& \multicolumn{1}{c|}{\multirow{1}{*}{~$m_{\tilde \chi_4^0}$~}}
& ($-387\thicksim-201$)$\cup$($219\thicksim 329$) \\
%%%\cline{2-6}
%\multicolumn{1}{|c|}{}
%& $110 \thicksim 160 $& \multicolumn{1}{|c|}{}& $105\thicksim 150$
%& \multicolumn{1}{|c|}{}& $800 \thicksim 2000 $\\
\cline{1-6}
\multicolumn{1}{c|}{\multirow{1}{*}{~$M_1$~}}& ($-236\thicksim-91$)$\cup$($115\thicksim 184$)
& \multicolumn{1}{c|}{\multirow{1}{*}{~$m_{A_1}$~}} & $251 \thicksim 569 $
& \multicolumn{1}{c|}{\multirow{1}{*}{~$m_{\tilde \chi_5^0}$~}} & ($-337\thicksim-238$)$\cup$($245\thicksim 1535$) \\
%%%\cline{2-6}
%\multicolumn{1}{|c|}{}& $-5000\thicksim 4500 $ & \multicolumn{1}{|c|}{}
%& $500 \thicksim 2050 $ & \multicolumn{1}{|c|}{} & $1100 \thicksim 3100 $ \\
\cline{1-6}
\multicolumn{1}{c|}{\multirow{1}{*}{~$M_2$~}}& $142\thicksim 1490 $
& \multicolumn{1}{c|}{\multirow{1}{*}{~$m_{A_2}$~}} & $1640 \thicksim 5190 $
& \multicolumn{1}{c|}{\multirow{1}{*}{~$m_{\tilde \chi_1^{\pm}}$~}} & $113 \thicksim 291 $ \\
%%%\cline{2-6}
%\multicolumn{1}{|c|}{}& $-5000\thicksim 4500 $ & \multicolumn{1}{|c|}{}
%& $500 \thicksim 2050 $ & \multicolumn{1}{|c|}{} & $1100 \thicksim 3100 $ \\
\cline{1-6}
\multicolumn{1}{c|}{\multirow{1}{*}{~$A_{\lambda}$~}}& ($-5000\thicksim-1500$)$\cup$($2070\thicksim 5000$)
& \multicolumn{1}{c|}{\multirow{1}{*}{~$m_{H^{\pm}}$~}} & $1640 \thicksim 5209 $
& \multicolumn{1}{c|}{\multirow{1}{*}{~$m_{\tilde \chi_2^{\pm}}$~}} & ($-1493\thicksim-220$)$\cup$($244\thicksim 1535$) \\
%%%\cline{2-6}
%\multicolumn{1}{|c|}{}& $-5000\thicksim 4500 $ & \multicolumn{1}{|c|}{}
%& $500 \thicksim 2050 $ & \multicolumn{1}{|c|}{} & $1100 \thicksim 3100 $ \\
\cline{1-6}
\multicolumn{1}{c|}{\multirow{1}{*}{~$A_{\kappa}$~}}& ($-593\thicksim-418$)$\cup$($271\thicksim 810$)
& \multicolumn{1}{c|}{\multirow{1}{*}{~$m_{\tilde l_L}$~}} & $119 \thicksim 1490 $
& \multicolumn{1}{c|}{\multirow{1}{*}{~$m_{\tilde l_R}$~}} & $100 \thicksim 1500 $ \\
%%%\cline{2-6}
%\multicolumn{1}{|c|}{}& $-5000\thicksim 4500 $ & \multicolumn{1}{|c|}{}
%& $500 \thicksim 2050 $ & \multicolumn{1}{|c|}{} & $1100 \thicksim 3100 $ \\
\cline{1-6}
\end{tabular}
}
\end{table}

%%%%%%%%%%%%%%%%%%%%%%%%%% F I G U R E %%%%%%%%%%%%%%%%%%%%%%%%%%%%%%%%%%%%%%%
\begin{figure}[t]
\begin{center}
\resizebox{0.7\textwidth}{!}{
\includegraphics[width=0.7\textwidth]{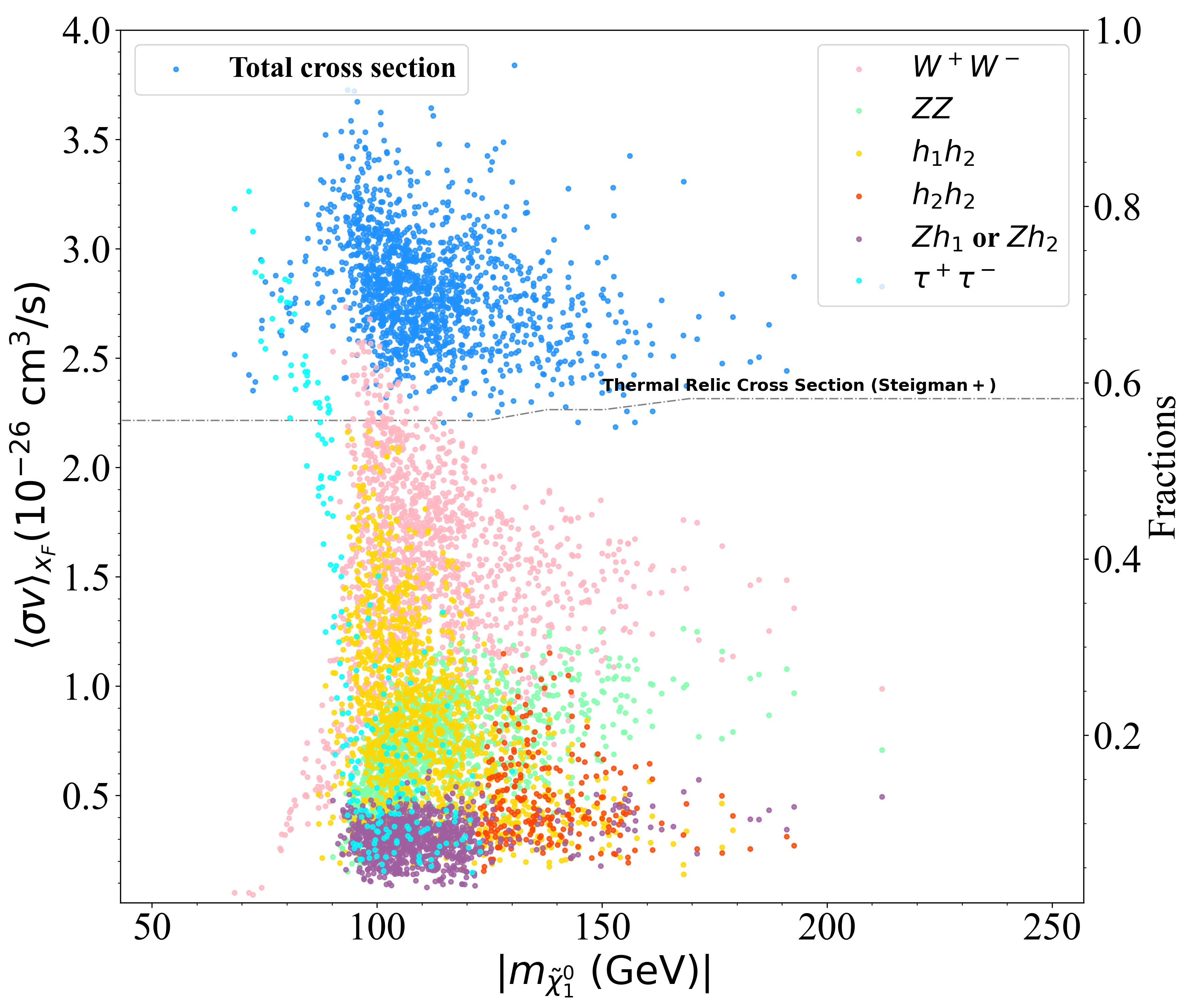}\hspace{-0.4cm}
}
\vspace{-0.4cm}
\caption{\label{fig5}The surviving samples that comply with the LZ-2022 limits are projected onto the $\langle \sigma v \rangle_{x_F}-m_{\tilde{\chi}_1^0}$ plane. Blue dots represent the total annihilation cross section (left axis), while pink, green, orange, red, purple, and cyan dots reprent the fraction of the annihilation channels $\tilde{\chi}_1^0 \tilde{\chi}_1^0 \to W^+ W^-, Z Z, h_1 h_2, h_2 h_2, Z h_1~or~Z h_2$, and $\tau^+ \tau^- $, respectively (right axis).
}
\end{center}
\end{figure}

Any DM candidate is also  subject to constraints from indirect detection experiments. In this study, we briefly examine the indirect detection limits on the surviving parameter points satisfied the LZ-2022 limits. Due to substantial astrophysical uncertainties in the production of charged particles, our analysis primarily focuses on constraints from photon emissions.

As illustrated in Fig.~\ref{fig5}, we analyze the thermally averaged annihilation cross section at freeze-out, $\left\langle \sigma v \right\rangle_{x_F}$, for the surviving samples. The dominant contribution arises from the $\tilde{\chi}_1^0 \tilde{\chi}_1^0 \to W^+ W^-$ annihilation channel, which mixes sizably with other annihilation channels including $\tilde{\chi}_1^0 \tilde{\chi}_1^0 \to ZZ$, $h_1 h_2$, $ h_2 h_2$, $ Z h_1$, and $ Z h_2$, to account for the DM relic density. These processes occur in the mass range $95~{\rm GeV} \lesssim m_{\tilde{\chi}_1^0} \lesssim 200~{\rm GeV}$. Additionally, for a few parameter points with $70~{\rm GeV} \lesssim m_{\tilde{\chi}_1^0} \lesssim 95~{\rm GeV}$, the  $\tilde {\chi}_1^0 \tilde {\chi}_1^0 \to \tau^+ \tau^- $ annihilation channel provides the dominant contribution.

The most stringent current limits on WIMP annihilation cross sections stem from Fermi-LAT and MAGIC observations of gamma rays from Milky Way satellite galaxies. These studies exclude the canonical thermal relic value $\langle \sigma v \rangle \sim 2 \times 10^{-26}\,{\rm cm}^3\,{\rm s}^{-1}$ for DM masses below approximately 100 GeV when annihilation predominantly produces $b$-quark or $\tau$-lepton final states~\cite{MAGIC:2016xys,Fermi-LAT:2015att,Fermi-LAT:2016uux}. For DM masses $m_{\tilde{\chi}_1^0} \gtrsim 100{\rm GeV} $, these constraints become weaker, constraining thermal annihilation cross sections larger than the canonical thermal relic value consistent with an acceptable relic density. Additional indirect detection bounds may arise from solar neutrino observations. DM particles captured by the Sun through elastic scattering can annihilate, producing neutrinos detectable at the Earth. In scenarios where annihilation primarily results in $W^+W^-$ final states---relevant to our model---current limits on the SD DM-proton scattering cross section are particularly stringent~\cite{Baum:2016oow,IceCube:2016yoy,Zornoza:2016xqb}. These analyses exclude $\sigma^{\text{SD}}_p \gtrsim 10^{-40}\text{cm}^2$ for DM masses up to $\sim 1\text{TeV}$.

In the viable regions of our model that are consistent with direct detection limits (both SI and SD) and  $\pm 10\%$ of the observed relic abundance, the dominant annihilation into $W^+W^-$ mixes sizably with other channels. Consequently, current indirect detections impose relatively weak limits on our parameter space.

\begin{table}[t]
\caption{The existing experimental searches for exotic Higgs decays $h\to ss\to XXYY$ at the 13 TeV LHC. Modified from Table 1 of Ref.~\cite{Carena:2022yvx}.}\label{table:XXYY}
\footnotesize\renewcommand\arraystretch{1.5}\centering
\begin{tabular}{|c|c|c|c|c|c|c|c|c|c|c|c|c|c|c|c|c|c|c|c|c|c|} \hline
Final state & Production mode & $m_s$ range [GeV] & $\mathcal{L}$ [fb$^{-1}$] & Collaboration \\ \hline
\multirow{2}{*}{$\mu\mu\mu\mu$} & \multirow{2}{*}{$gg$ fusion} & $[4,8]\cup[11.5,60]$ & 137 & CMS~\cite{CMS:2021pcy} \\ \cline{3-5}
 & & $[1.2,2]\cup[4.4,8]\cup[12,60]$ & 139 & ATLAS~\cite{ATLAS:2021ldb} \\ \hline
\multirow{1}{*}{$\mu\mu\tau\tau$} & \multirow{1}{*}{$gg$ fusion} & $[15,62.5]$ & 35.9 & CMS~\cite{CMS:2018qvj} \\ \hline
\multirow{2}{*}{$bb\mu\mu$} & \multirow{2}{*}{$gg$ fusion} & $[18,60]$ & 139 & ATLAS~\cite{ATLAS:2021hbr} \\ \cline{3-5}
& & $[20,62.5]$ & 35.9 & CMS~\cite{CMS:2018nsh} \\ \hline
$bb\tau\tau$ & $gg$ fusion & $[15,60]$ & 35.9 & CMS~\cite{CMS:2018zvv} \\ \hline
\multirow{1}{*}{$bbbb$} & $Wh/Zh$ & $[20,60]$ & 36.1 & ATLAS~\cite{ATLAS:2020ahi} \\ \hline
$\gamma\gamma\gamma\gamma$ & $gg$ fusion & $[15,60]$ & 132 & CMS~\cite{CMS:2021bvh} \\ \hline
$\gamma\gamma jj$ & VBF & $[20,60]$ & 36.7 & ATLAS~\cite{ATLAS:2018jnf} \\ \hline
\end{tabular}
\end{table}

%%%%%%%%%%%%%%%%%%%%%%%%%% F I G U R E %%%%%%%%%%%%%%%%%%%%%%%%%%%%%%%%%%%%%%%
\begin{figure}[t]
\begin{center}
\resizebox{1.\textwidth}{!}{
\includegraphics[width=0.9\textwidth]{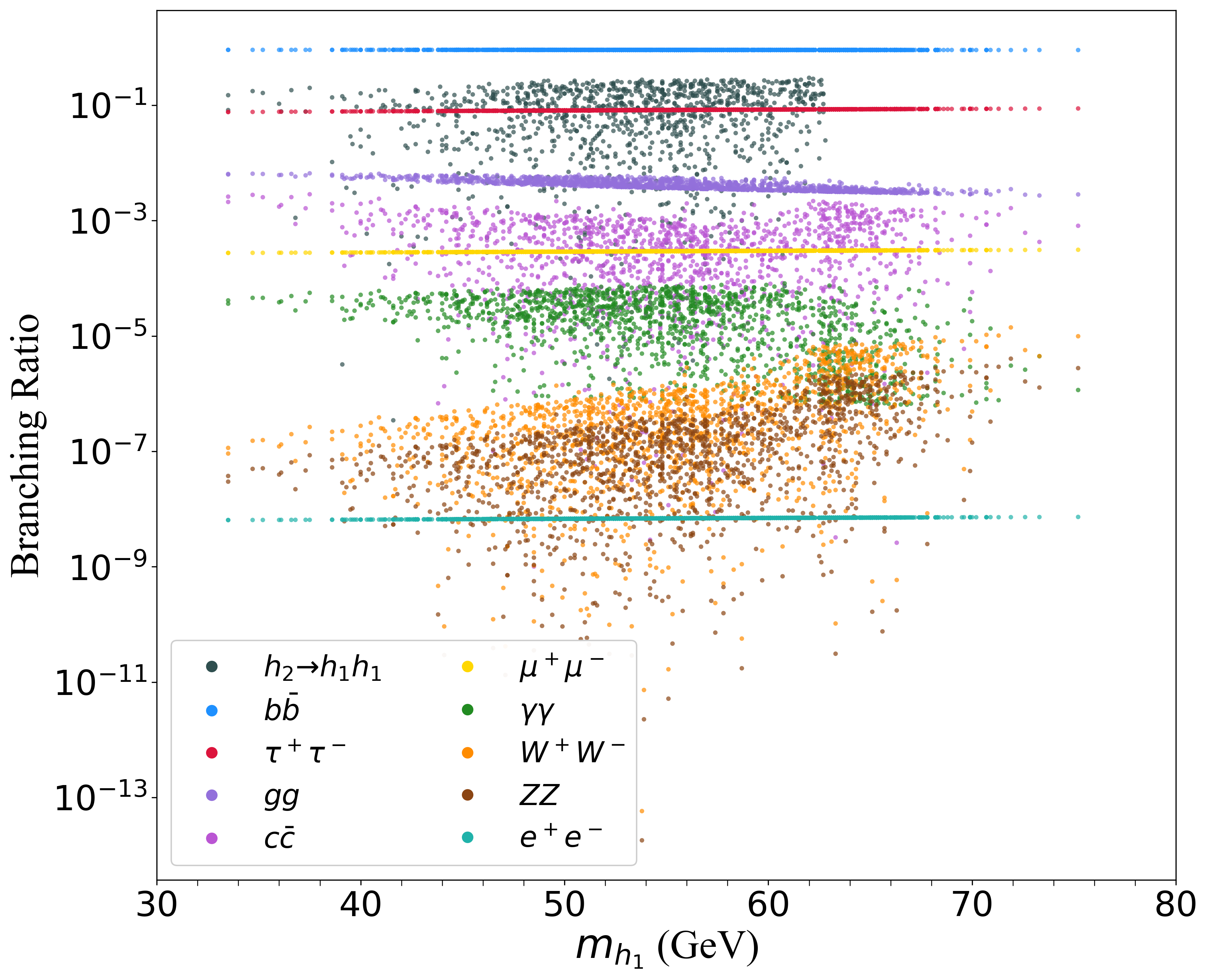}\hspace{0.1cm}
\includegraphics[width=0.9\textwidth]{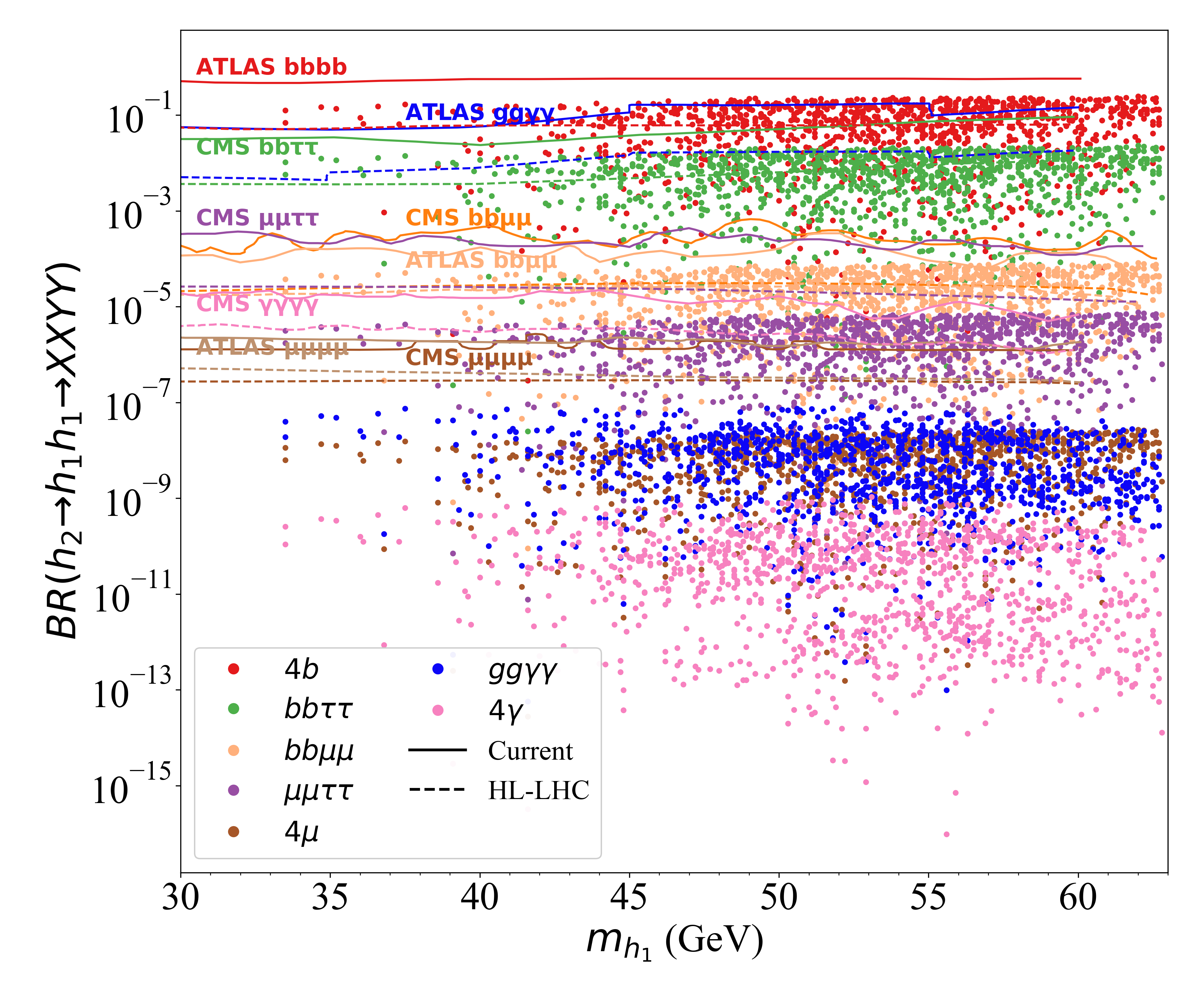}
}
\vspace{-0.8cm}
\caption{\label{fig6}The left figure shows the BRs for $h_2 \to h_1 h_1$ as well as the BRs of $h_1$ into SM particles. The right figure illustrates the BRs for $h_2 \to h_1 h_1 \to XXYY$, current experimental bounds (solid lines), and corresponding projections at the high-luminosity LHC (HL-LHC) (dashed lines). All points comply with the LZ-2022 limits. }
\end{center}
\end{figure}

\subsection{\label{Collider}Collider constraints}
\begin{table}[t]
	\caption{Experimental analyses included in the package \texttt{SModelS-3.0.0}.}
	\label{tab:SModelS}
	\vspace{0.1cm}
	\resizebox{1.0 \textwidth}{!}{
		\begin{tabular}{cccc}
			\hline\hline
			\texttt{Name} & \texttt{Scenario} &\texttt{Final State} &$\texttt{Luminosity} (\texttt{fb}^{\texttt{-1}})$ \\\hline
 			\begin{tabular}[l]{@{}l@{}}ATLAS-1402-7029~\cite{ATLAS:2014ikz}\end{tabular} & \begin{tabular}[c]{@{}c@{}c@{}c@{}} $\tilde{\chi}_1^{\pm}\tilde{\chi}_{2}^0\rightarrow \tilde{\ell}\nu\tilde{\ell}\ell(\tilde{\nu}\nu), \ell\tilde{\nu}\tilde{\ell}\ell(\tilde{\nu}\nu)\rightarrow \ell\nu\tilde{\chi}_1^0\ell\ell(\nu\nu)\tilde{\chi}_1^0$ \\ $\tilde{\chi}_1^{\pm}\tilde{\chi}_{2}^0\rightarrow\tilde\tau\nu\tilde\tau\tau(\tilde\nu\nu), \tau\tilde\nu\tilde\tau\tau(\tilde\nu\nu)\rightarrow\tau\nu\tilde{\chi}_1^0\tau\tau(\nu\nu)\tilde{\chi}_1^0$ \\ $\tilde{\chi}_1^{\pm}\tilde{\chi}_{2}^0\rightarrow W^\pm\tilde{\chi}_1^0Z\tilde{\chi}_1^0\rightarrow \ell\nu\tilde{\chi}_1^0\ell\ell\tilde{\chi}_1^0$ \\$\tilde{\chi}_1^{\pm}\tilde{\chi}_{2}^0\rightarrow W^\pm\tilde{\chi}_1^0h\tilde{\chi}_1^0\rightarrow \ell\nu\tilde{\chi}_1^0\ell\ell\tilde{\chi}_1^0$ \end{tabular} & 3$ \ell$ + $ E_{\rm T}^{\rm miss}$ & 20.3 \\ \\
 \begin{tabular}[l]{@{}l@{}}CMS-SUS-16-034~\cite{CMS:2017kxn}\end{tabular}&$\tilde{\chi}_2^0\tilde{\chi}_1^{\pm}\rightarrow W\tilde{\chi}_1^0Z(h)\tilde{\chi}_1^0$ & n$\ell$(n\textgreater{}=2) + nj(n\textgreater{}=1) + $E_{\rm T}^{\rm miss}$ & 35.9 \\ \\
			\begin{tabular}[l]{@{}l@{}} CMS-SUS-16-039~\cite{CMS:2017moi} \end{tabular} &\begin{tabular}[c]{@{}c@{}c@{}c@{}c@{}} $\tilde{\chi}_2^0\tilde{\chi}_1^{\pm}\rightarrow \ell\tilde{\nu}\ell\tilde{\ell}$\\$\tilde{\chi}_2^0\tilde{\chi}_1^{\pm}\rightarrow\tilde{\tau}\nu\tilde{\ell}\ell$\\$\tilde{\chi}_2^0\tilde{\chi}_1^{\pm}\rightarrow\tilde{\tau}\nu\tilde{\tau}\tau$\\ $\tilde{\chi}_2^0\tilde{\chi}_1^{\pm}\rightarrow WZ\tilde{\chi}_1^0\tilde{\chi}_1^0$\\$\tilde{\chi}_2^0\tilde{\chi}_1^{\pm}\rightarrow WH\tilde{\chi}_1^0\tilde{\chi}_1^0$\end{tabular} & n$\ell(n\textgreater{}0)$($\tau$) + $E_{\rm T}^{\rm miss}$& 35.9\\ \\			
 \begin{tabular}[l]{@{}l@{}}CMS-SUS-16-045~\cite{CMS:2017bki}\end{tabular} &$\tilde{\chi}_2^0\tilde{\chi}_1^{\pm}\rightarrow W^{\pm}\tilde{\chi}_1^0h\tilde{\chi}_1^0$& 1$ \ell$ 2b + $ E_{\rm T}^{\rm miss}$ & 35.9 \\ \\
 			\begin{tabular}[l]{@{}l@{}}CMS-SUS-16-048~\cite{CMS:2018kag}\end{tabular}&\begin{tabular}[c]{@{}c@{}c@{}}$\tilde{\chi}_2^0\tilde{\chi}_1^{\pm}\rightarrow Z\tilde{\chi}_1^0W\tilde{\chi}_1^0$\\ $\tilde{\chi}_2^0\tilde{\chi}_1^0\rightarrow Z\tilde{\chi}_1^0\tilde{\chi}_1^0$\end{tabular}& 2$\ell$ + $E_{\rm T}^{\rm miss}$ & 35.9 \\ \\
			\begin{tabular}[l]{@{}l@{}} CMS-SUS-17-004~\cite{CMS:2018szt}\end{tabular} &$\tilde{\chi}_{2}^0\tilde{\chi}_1^{\pm}\rightarrow Wh(Z)\tilde{\chi}_1^0\tilde{\chi}_1^0$ & n$ \ell$(n\textgreater{}=0) + nj(n\textgreater{}=0) + $ E_{\rm T}^{\rm miss}$ & 35.9 \\ \\			\begin{tabular}[l]{@{}l@{}} CMS-SUS-17-009~\cite{CMS:2018eqb}\end{tabular} &$\tilde{\ell}\tilde{\ell}\rightarrow \ell\tilde{\chi}_1^0\ell\tilde{\chi}_1^0$ &2$\ell$ + $E_{\rm T}^{\rm miss}$ & 35.9 \\ \\			\begin{tabular}[l]{@{}l@{}} CMS-SUS-17-010~\cite{CMS:2018xqw}\end{tabular} &\begin{tabular}[c]{@{}c@{}}$\tilde{\chi}_1^{\pm}\tilde{\chi}_1^{\mp}\rightarrow W^{\pm}\tilde{\chi}_1^0 W^{\mp}\tilde{\chi}_1^0$\\$\tilde{\chi}_1^{\pm}\tilde{\chi}_1^{\mp}\rightarrow \nu\tilde{\ell} \ell\tilde{\nu}$ \\ \end{tabular}&2$ \ell$ + $E_{\rm T}^{\rm miss}$ & 35.9 \\ \\
 			\begin{tabular}[l]{@{}l@{}}ATLAS-1803-02762~\cite{ATLAS:2018ojr}\end{tabular} &\begin{tabular}[c]{@{}c@{}c@{}c@{}}$\tilde{\chi}_2^0\tilde{\chi}_1^{\pm}\rightarrow WZ\tilde{\chi}_1^0\tilde{\chi}_1^0$\\$\tilde{\chi}_2^0\tilde{\chi}_1^{\pm}\rightarrow \nu\tilde{\ell}l\tilde{\ell}$\\$\tilde{\chi}_1^{\pm}\tilde{\chi}_1^{\mp}\rightarrow \nu\tilde{\ell}\nu\tilde{\ell}$\\ $ \tilde{\ell}\tilde{\ell}\rightarrow \ell\tilde{\chi}_1^0\ell\tilde{\chi}_1^0$\end{tabular} & n$ \ell$ (n\textgreater{}=2) + $ E_{\rm T}^{\rm miss}$ & 36.1 \\ \\
 			\begin{tabular}[l]{@{}l@{}}ATLAS-1806-02293~\cite{ATLAS:2018eui}\end{tabular} &$\tilde{\chi}_2^0\tilde{\chi}_1^{\pm}\rightarrow WZ\tilde{\chi}_1^0\tilde{\chi}_1^0$ &n$\ell$(n\textgreater{}=2) + nj(n\textgreater{}=0) + $ E_T^{miss}$ & 36.1 \\ \\
			\begin{tabular}[l]{@{}l@{}}ATLAS-1812-09432~\cite{ATLAS:2018qmw}\end{tabular} &$\tilde{\chi}_2^0\tilde{\chi}_1^{\pm}\rightarrow Wh\tilde{\chi}_1^0\tilde{\chi}_1^0$ & n$ \ell$ (n\textgreater{}=0) + nj(n\textgreater{}=0) + nb(n\textgreater{}=0) + n$\gamma$(n\textgreater{}=0) + $E_{\rm T}^{\rm miss}$ & 36.1 \\ \\
 \begin{tabular}[l]{@{}l@{}}CMS-SUS-20-001~\cite{CMS:2020bfa}\end{tabular}&\begin{tabular}[c]{@{}c@{}c@{}}$\tilde{\chi}_2^0\tilde{\chi}_1^{\pm}\rightarrow Z\tilde{\chi}_1^0W\tilde{\chi}_1^0$\\ $\tilde{\ell}\tilde{\ell}\rightarrow \ell\tilde{\chi}_1^0\ell\tilde{\chi}_1^0$\end{tabular}& 2$\ell$ + $E_{\rm T}^{\rm miss}$ & 137 \\ \\
 \begin{tabular}[l]{@{}l@{}}CMS-SUS-20-004~\cite{CMS:2022vpy}\end{tabular}&$\tilde{\chi}_2^0\tilde{\chi}_3^{0}\rightarrow h\tilde{\chi}_1^0h\tilde{\chi}_1^0$ & 4b + $E_{\rm T}^{\rm miss}$ & 137 \\ \\
 \begin{tabular}[l]{@{}l@{}}CMS-SUS-21-002~\cite{CMS:2022sfi}\end{tabular}&\begin{tabular}[c]{@{}c@{}c@{}c@{}}$\tilde{\chi}_1^{\pm}\tilde{\chi}_1^{\mp}\rightarrow W^{\pm}\tilde{\chi}_1^0 W^{\mp}\tilde{\chi}_1^0$ \\ $\tilde{\chi}_{2/3}^0\tilde{\chi}_1^{\pm}\rightarrow W^{\pm}\tilde{\chi}_1^0 Z(h)\tilde{\chi}_1^0$ \\ $\tilde{\chi}_2^0\tilde{\chi}_3^{0}\rightarrow Z\tilde{\chi}_1^0h\tilde{\chi}_1^0$ \end{tabular}& 4b + $E_{\rm T}^{\rm miss}$ & 137 \\ \\
			\begin{tabular}[l]{@{}l@{}}ATLAS-1908-08215~\cite{ATLAS:2019lff}\end{tabular} &\begin{tabular}[c]{@{}c@{}}$\tilde{\ell}\tilde{\ell}\rightarrow \ell\tilde{\chi}_1^0\ell\tilde{\chi}_1^0$\\$\tilde{\chi}_1^{\pm}\tilde{\chi}_1^{\mp}$ \\ \end{tabular} & 2$\ell$ + $ E_{\rm T}^{\rm miss}$ & 139 \\ \\
			\begin{tabular}[l]{@{}l@{}}ATLAS-1909-09226~\cite{ATLAS:2020pgy}\end{tabular} & $\tilde{\chi}_{2}^0\tilde{\chi}_1^{\pm}\rightarrow Wh\tilde{\chi}_1^0\tilde{\chi}_1^0$ & 1$ \ell$ + h($\bm \rightarrow$ bb) + $ E_{\rm T}^{\rm miss}$ & 139 \\ \\
 			\begin{tabular}[l]{@{}l@{}}ATLAS-1911-12606~\cite{ATLAS:2019lng}\end{tabular} & \begin{tabular}[c]{@{}c@{}c@{}c@{}} $\tilde{\chi}_{2}^0\tilde{\chi}_1^{\pm}\rightarrow Z\tilde{\chi}_1^0W\tilde{\chi}_1^0$ \\ $\tilde{\chi}_{2}^0\tilde{\chi}_1^{0}\rightarrow Z\tilde{\chi}_1^0\tilde{\chi}_1^0$ \\ $\tilde{\chi}_{1}^+\tilde{\chi}_1^-\rightarrow W^+\tilde{\chi}_1^0W^-\tilde{\chi}_1^0$ \\ $\tilde{\ell}\tilde{\ell}\rightarrow \ell\tilde{\chi}_1^0\ell\tilde{\chi}_1^0$\end{tabular} & 2$ \ell$ + 2j + $ E_{\rm T}^{\rm miss}$ & 139 \\ \\
 		\begin{tabular}[l]{@{}l@{}}ATLAS-1912-08479~\cite{ATLAS:2019wgx}\end{tabular} &$\tilde{\chi}_2^0\tilde{\chi}_1^{\pm}\rightarrow W(\rightarrow l\nu)\tilde{\chi}_1^0Z(\rightarrow\ell\ell)\tilde{\chi}_1^0$& 3$\ell $ + $ E_{\rm T}^{\rm miss}$ & 139 \\ \\
 	\begin{tabular}[l]{@{}l@{}}ATLAS-2106-01676~\cite{ATLAS:2021moa}\end{tabular} & \begin{tabular}[c]{@{}c@{}}$\tilde{\chi}_1^{\pm}\tilde{\chi}_{2}^0\rightarrow W(\rightarrow \ell \nu)Z(\rightarrow \ell\ell)\tilde{\chi}_1^0\tilde{\chi}_1^0$\\ $\tilde{\chi}_1^{\pm}\tilde{\chi}_{2}^0\rightarrow W(\rightarrow \ell \nu)h(\rightarrow \ell\ell)\tilde{\chi}_1^0\tilde{\chi}_1^0$ \end{tabular} & 3$ \ell$ + $ E_{\rm T}^{\rm miss}$ & 139 \\ \\
 			\begin{tabular}[l]{@{}l@{}}ATLAS-2108-07586~\cite{ATLAS:2021yqv}\end{tabular} & \begin{tabular}[c]{@{}c@{}}$\tilde{\chi}_1^{\pm}\tilde{\chi}_1^{\pm}\rightarrow WW\tilde{\chi}_1^0\tilde{\chi}_1^0$ \\ $\tilde{\chi}_1^{\pm}\tilde{\chi}_{2}^0\rightarrow WZ(h)\tilde{\chi}_1^0\tilde{\chi}_1^0$ \end{tabular}& 4j + $ E_{\rm T}^{\rm miss}$ & 139 \\ \\
 		\begin{tabular}[l]{@{}l@{}}ATLAS-2204-13072~\cite{ATLAS:2022zwa}\end{tabular} & $\tilde{\chi}_{2}^0\tilde{\chi}_1^{\pm}\rightarrow W(\rightarrow q q
 )\tilde{\chi}_1^0Z(\rightarrow \ell\ell)\tilde{\chi}_1^0$ & 2$ \ell$ + 2j + $ E_{\rm T}^{\rm miss}$ & 139 \\ \\
 			\begin{tabular}[l]{@{}l@{}}ATLAS-2209-13935~\cite{ATLAS:2022hbt}\end{tabular} & \begin{tabular}[c]{@{}c@{}}$\tilde{\ell}\tilde{\ell}\rightarrow \ell\ell\tilde{\chi}_1^0\tilde{\chi}_1^0$ \\ $\tilde{\chi}_{1}^{\pm}\tilde{\chi}_1^{\mp}\rightarrow W(\rightarrow \ell \nu)W(\rightarrow \ell \nu)\tilde{\chi}_1^0\tilde{\chi}_1^0$ \end{tabular} & 2$ \ell$ + $E_{\rm T}^{\rm miss}$ & 139 \\ \hline\\
	\end{tabular}}
\end{table}

The findings from the previous section indicate that light neutralino DM exhibits a limited yet viable survival potential under the constraints of the LZ-2022 experiment. Building upon this, we examined the collider phenomenology associated with these viable scenarios.

Firstly, we examine the constraints imposed by the invisible decay modes of the SM-like Higgs boson. As shown in Table~\ref{table-1}, the mass range of the lightest Higgs boson ($h_1$) lies within the range $(30,75)~{\rm GeV}$. For $30 ~{\rm GeV}<m_{h_1}<m_h/2\approx 63~{\rm GeV}$, the decay channel $h_2 \to h_1h_1$ becomes kinematically accessible. This mass range can be further extended up to about 75 GeV when three-body decays are taken into account. As $m_{h_1}$ increases, the decay mode becomes progressively off-shell, thereby reducing its overall contribution. The decay $h_2 \to h_1h_1$ leads to a variety of final states, depending on the subsequent decays of $h_1$. These can be generically represented as XXYY, where X and Y denote SM particles. For $m_{h_1}>2m_b\sim10$ GeV, the dominant decay channel is $h_2 \to h_1h_1 \to bbbb$, whereas for $m_{h_1}<10$ GeV, the final states include $gggg$, $gg\tau\tau$, and $\tau\tau\tau\tau$~\cite{Gershtein:2020mwi,Spira:1997dg,Winkler:2018qyg}. Table~\ref{table:XXYY} summarizes relevant LHC searches targeting $h \to ss \to XXYY$, where the intermediate scalar particle $s$ has a mass in the range $30-63~{\rm GeV}$. The table outlines the final states investigated, the production mechanisms, integrated luminosities, and the scalar $s $ mass ranges considered. Notably, for all final states apart from \( bbbb \), the LHC analyses require at least one pair of non-hadronic final-state particles, such as photons, muons, or $\tau$ leptons.

All points complying with the LZ-2022 limits are projected in Fig.~\ref{fig6}. The left panel of Fig.~\ref{fig6} shows the BRs for the exotic decay $ h_2 \to h_1h_1 $ as well as the BRs of the CP-even singlet Higgs $ h_1 $ into SM particles. The right panel displays the combined BRs for $h_2 \to h_1 h_1 \to XXYY$, superimposed with current experimental exclusion limits (solid lines) and projected HL-LHC sensitivities (dashed lines). The HL-LHC  sensitivities are derived by assuming all uncertainties scale with $1/\sqrt{L}$~\cite{Carena:2022yvx}. Regions above the exclusion curves are ruled out by present experimental data.
All $XXYY$ channels corresponding to the surviving samples remain consistent with current bounds. Even under HL-LHC projections, channels such as $bbbb$, $bb\tau\tau$, and $bb\mu\mu$ continue to exhibit viable parameter space. In addition, final states including $\mu\mu\tau\tau$, $\mu\mu\mu\mu$, $gg\gamma\gamma$, and $\gamma\gamma\gamma\gamma$ fully satisfy all existing constraints.

Next, we assessed whether the surviving parameter samples comply with the constraints from LHC EWinos searches using the \texttt{SModelS-3.0.0} program~\cite{Khosa:2020zar}, which includes the experimental selection efficiencies listed in Table~\ref{tab:SModelS}. Our analysis shows that the most stringent constraints arise from the direct EWino searches in the $WZ$ mediated $2l+ E_{\rm T}^{\rm miss}$ channel~\cite{CMS:2020bfa} and the $Wh_{125}$ mediated $1l+2b+E_{\rm T}^{\rm miss}$ channel~\cite{ATLAS:2020pgy} implemented in \texttt{SModels-3.0.0}. The parameter samples that pass the LHC EWinos search constraints fall within the mass range $140 {\rm GeV} < \tilde{\chi}^0_2 \approx \tilde{\chi}^{\pm}_1 < 260 {\rm GeV}$ and mass splitting $45 {\rm GeV} < \Delta m = \tilde{\chi}^0_2 - \tilde{\chi}^0_1 < 65 {\rm GeV}$.  After applying constraints from the LHC supersymmetry search, the sample size decreased from 2924 to 401, highlighting the substantial impact of LHC constraints on this type of spectrum.

%Recent searches for the ``golden channel'',
%$pp \to \tilde{\chi}^0_2 \tilde{\chi}^{\pm}_1 \to \tilde{\chi}^0_1 Z^{(*)} \, \tilde{\chi}^0_1 W^{\pm (*)}$ show consistent excesses between ATLAS and CMS in the 2~lepton, 3-lepton and mono-jet
%searches, assuming $\tilde{\chi}^0_2 \approx \tilde{\chi}^{\pm}_1 > 200 GeV$ and
%$\Delta m = \tilde{\chi}^0_2 - \tilde{\chi}^0_1 \approx 20 GeV$. SUSY scenarios with respect to the observed excesses, taking into account all relevant experimental constraints, has recently been highlighted~\cite{Chakraborti:2024pdn,Agin:2024yfs}. While each search and experiment individually is not significant by itself, the occurence of excesses in multiple search channels, observed by both ATLAS and CMS, gives rise to
%the hope that finally a glimpse of BSM physics has been observed. We are eagerly awaiting the corresponding upcoming Run~3 results.
%%%%%%%%%%%%%%%%%%%%%%%%% F I G U R E %%%%%%%%%%%%%%%%%%%%%%%%%%%%%%%%%%%%%%%
\begin{figure}[h!]
\begin{center}
\includegraphics[width=0.9\textwidth]{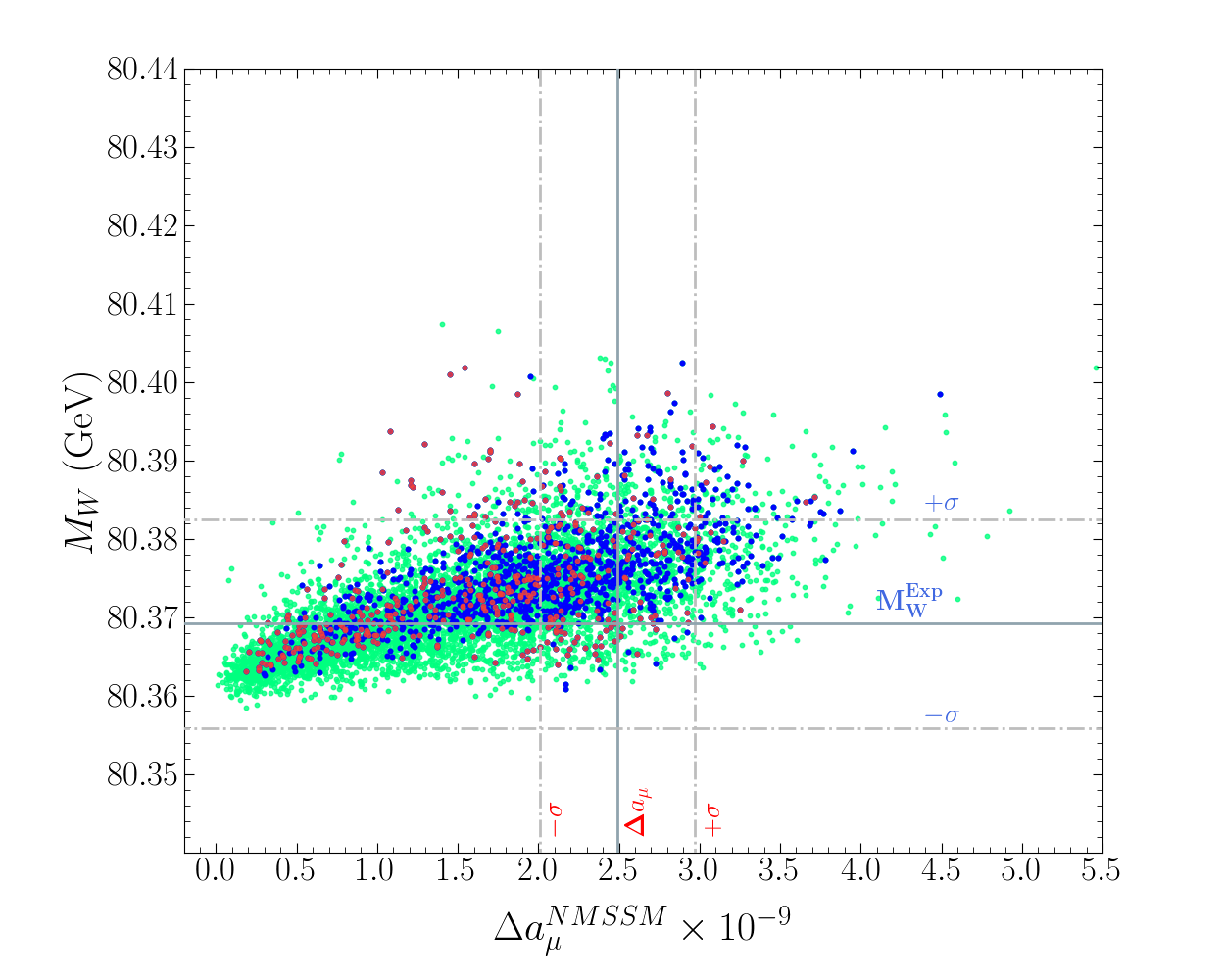}
\caption{\label{fig:amu-mw}
Results for the surviving scenarios in the $\Delta a_{\mu}- M_W$ plane. Green points represent all the refined samples, blue points denote the samples satisfying the LZ-2022 limits, and red points correspond to those that comply with both the LZ-2022 limits and the constraints from direct EWinos searches as implemented in \texttt{SModelS-3.0.0}.
The vertical grey lines indicate the central value of $\Delta a_{\mu} $ as given in Eq.~(\ref{eq:muong-2}) (solid line) and its $\pm 1\,\sigma$ range (dashed lines).
The horizontal grey lines show the current central value for the world average without CDF II of the W boson mass $M_W^{exp}$
(solid line), and its $\pm 1\,\sigma$ uncertainties (dashed lines).
}
\end{center}
\vspace{-1em}
\end{figure}
%%%%%%%%%%%%%%%%%%%%%%%%%% F I G U R E %%%%%%%%%%%%%%%%%%%%%%%%%%%%%%%%%%%%%%% %%%	

Finally, we mapped the results onto the $\Delta a_{\mu}- M_W$ plane in Fig.\ref{fig:amu-mw}, where the green, blue, and red points indicate all the refined samples, the  samples that comply with the LZ-2022 limits, and the samples that satisfy both the LZ-2022 limits and the limits from direct EWinos searches in \texttt{SModelS-3.0.0} program, respectively. The vertical grey lines denote the  central value of $\Delta a_{\mu} $, as given in Eq.~(\ref{eq:muong-2}) (solid line), along with its $\pm 1\, \sigma$  range (dashed lines). The horizontal grey lines represent the current central value for the world average of the W boson mass  $M_W^{Exp}$
(solid line), and its $\pm 1\,\sigma$ uncertainties (dashed lines), excluding CDF II data~\cite{LHC-TeVMWWorkingGroup:2023zkn,ParticleDataGroup:2024cfk}~\footnote{A recent combination of $W$-boson mass measurements performed by the ATLAS, LHCb, CDF, and D0 experiments, using precise calibration of simulation tools and PDFs, has produced the world average, $M_W = 80.3946 \pm 0.0115$ GeV ~\cite{LHC-TeVMWWorkingGroup:2023zkn}. However, the probability of compatibility is $0.5\%$ or less, depending on the selected PDF set, largely due to the W mass measurement by CDF from Run II at the Tevatron ~\cite{LHC-TeVMWWorkingGroup:2023zkn}, $M_W = 80.432 \pm 0.016$ GeV (adjusted to the common PDF set CT18~\cite{Hou:2019qau} in Ref.~\cite{LHC-TeVMWWorkingGroup:2023zkn}). This value differs by almost $3.6\sigma$ from the other measurements of $M_W$, while the latter agree well among themselves, with the average excluding CDF II being $M_W = 80.3692 \pm 0.0133 $ GeV~\cite{LHC-TeVMWWorkingGroup:2023zkn}.}. Interestingly, the particle spectra obtained through scanning effectively describe both $M_W$ and $a_\mu$. It is evident that the predicted values of $M_W$ for the majority of the refined samples lie within the $1\,\sigma$ confidence interval of $M_W^{Exp}$. A small fraction of the samples shows predicted values exceeding the upper bound of this confidence interval. This result can be attributed to the presence of light EWinos in the 100~GeV$-$1~TeV range, where electroweakly-interacting SUSY particles contribute significantly to $M_W$. Additionally, for lower $\Delta a_{\mu}$ values, the reduced $M_W^{NMSSM}$ values suggest heavier electroweak supersymmetric particle masses. Therefore, the relatively light SUSY particles necessary for larger $\Delta a_{\mu}$ give rise to a slight increase in the prediction for $M_W$, which is independent of the variation of the other parameters in the scan. These conclusions are similar to the studies in Ref.~\cite{Domingo:2022pde,Tang:2022pxh,Bagnaschi:2022qhb}.

\section{Conclusion}\label{sec:sum}

In light of recent advancements in particle physics experiments, we have explored the implications of these developments on the $\mathbb{Z}_3$-NMSSM featuring  a light bino-dominated LSP and a light singlet-like scalar. Our study specifically examines the impact of the muon g-2 measurement at Fermilab, the LHC SUSY search, and the DM direct detection by the LZ-2022 experiment, as these experiments are sensitive to different parameters and provide complementary insights into the $\mathbb{Z}_3$-NMSSM.

 We began our analysis by utilizing the MultiNest algorithm to scan a broad parameter space, guided by LZ-2022 experiments data, constraints from the LHC Higgs date, the muon g-2 observable, and B-physics measurements.  Based on the composition of
 $ {\tilde{\chi}_1^0}$, the surviving samples were categorized into two groups: bino-higgsino ($ N_{15}^2 < 0.05 $) and bino-higgsino-singlino ($N_{15}^2 > 0.05$). These systems exhibit annihilation channels into \( W^+ W^- \), \( Z Z \), \( h_1 h_2 \), \( h_2 h_2 \), \( Z h_1 \), and \( Z h_2 \) to satisfy relic density requirements, facilitated by both a non-negligible higgsino component in the LSP and the presence of a light singlet-like scalar. However, their viability is challenged by stringent DM direct detection constraints. The bino-higgsino system’s $\sigma^{SI}$ can be below the LZ-2022 experimental limit, as well as the neutrino floor under blind-spot conditions. However, its $\sigma^{SD}$ is significantly constrained by the LZ experiment due to large $|N_{13}^2-N_{14}^2|$ values, which resembles the MSSM scenario with a higgsino-like NLSP ~\cite{Dutta:2014hma}. This results in the requirement of a larger $\mu_{eff}$ under the LZ-2022 experiment~\cite{He:2023lgi,Li:2023hey}. In contrast, for the bino-higgsino-singlino system, the values of $|N_{13}^2-N_{14}^2|$ approach zero corresponding to the blind-spot condition for the SD scattering cross section, which results in $\sigma_{\tilde{\chi}_1^0-n}^{SD}$ as low as $10^{-50}$ cm$^2$. However, this leads to a significant bino-singlino mixing in the LSP, which may be disfavored by constraints on the SI cross section. A moderate to large value of `$\lambda$' ($0.3 < \lambda < 0.7$) plays a crucial role in adequately tempering the bino LSP with higgsino (and hence singlino) admixture. It is important to note that in the bino-higgsino-singlino system, the SI cross sections for proton and neutron are abnormal below the order of $10^{-46} {\rm cm^2} $, that is, when one value is significantly higher, its counterpart may be relatively lower. This may be related to the small difference between $F_{q=u,d} ^ {(p)}$ and $F_{q=u,d} ^ {(n)}$. In this case, we utilize the effective cross section when considering the restriction of DM direct detection constraints on SI cross section.

 Exotic Higgs decays are a key aspect of discovery programs at both current and future colliders \cite{Curtin:2013fra,FCC:2018byv,CEPCStudyGroup:2018ghi,FCC:2018vvp,FCC:2018evy,Cepeda:2021rql}. For the surviving samples that comply with the LZ-2022 limits, the mass range for the lightest Higgs boson ($h_1$) is between $(30,75)~{\rm GeV}$. When $30 ~{\rm GeV}<m_{h_1}<m_h/2\approx 63~{\rm GeV}$, the decay channel $h_2 \to h_1h_1$ is kinematically possible. Therefore, we explore the constraints imposed by the invisible decay of the SM-like
Higgs boson. A dedicated search for exotic Higgs decays could effectively probe this framework at the LHC, while future exotic Higgs decay searches at the high-luminosity LHC and upcoming Higgs factories will be critical for thoroughly investigating this scenario. At the HL-LHC, detector upgrades, new trigger and analysis strategies, and larger datasets will progressively enhance sensitivity to small BRs, particularly in cleaner, subdominant final states (e.g., $h\to ss\to bb \tau \tau$, $h\to$ invisible) \cite{LHCHiggsCrossSectionWorkingGroup:2016ypw,Cepeda:2019klc}. Meanwhile, proposed $e^+ e^-$ colliders offer lower integrated luminosity but much lower backgrounds, providing excellent sensitivity to challenging all-hadronic modes, such as $h\to ss\to bbbb$ \cite{Liu:2016zki}. Additionally, the surviving scenario is characterized by $M_{1} \in (-291,-35){~\rm GeV}~ \bigcup~ (36,293){~\rm GeV}$, $\mu_{eff} \in (-311,-101){~\rm GeV}~ \bigcup~ (102,305){~\rm GeV}$, $m_{\tilde{S}} \in (-687,-125){~\rm GeV}~ \bigcup~ (112,564){~\rm GeV}$, $M_{2} \in (114,1500){~\rm GeV}$, $|\mu_{eff}/M_1| \in (1.2,4.53) $, $|m_{\tilde{S}}/\mu_{eff}| \in (1.0,3.9) $, $|m_{\tilde{\chi}_1^0}/M_1| \in (0.71,0.98) $. $|m_{\tilde{\chi}_2^0}/\mu_{eff}| \in (0.7,1.13) $, $|m_{\tilde{\chi}_3^0}/\mu_{eff}| \in (1.0,1.33) $, $|m_{\tilde{\chi}_4^0}/m_{\tilde{S}}| \in (0.7,1.5)$, and $ |m_{\tilde{\chi}_1^{\pm}}/\mu_{eff}|\in (0.7,1.0)$. These parameters are tightly constrained by the direct searches of EWinos at the LHC in the modes
$pp \to m_{\tilde{\chi}_1^{\pm}} m_{\tilde{\chi}_2^{0}} \to m_{\tilde{\chi}_1^{0}} W^{\pm (*)}, m_{\tilde{\chi}_1^{0}} Z^{(*)}/h_{SM} $ but can simultaneously explain the $ M_W$ and $a_\mu$ anomaly.

In conclusion, the $\mathbb{Z}_3$-NMSSM, featuring a light bino-dominated LSP and a light singlet-like scalar, faces significant challenges due to increasingly stringent upper limits on both SI and SD DM-nucleus elastic scattering cross sections, along with constraints from LHC EWino searches. However, this scenario still has the potential to naturally accommodate the observed $Z$ boson mass, reproduce the SM-like Higgs boson mass, account for the muon anomalous magnetic moment, and contribute to the $W$ boson mass. These phenomena are characteristic of the NMSSM, wherein the bino-dominated LSP is tempered by the singlino, with assistance from the higgsinos.

\section*{Acknowledgments}
We sincerely thank Prof. Junjie Cao for numerous helpful discussions. We thank LetPub (www.letpub.com.cn) for its linguistic assistance during the preparation of this manuscript. This work is supported by  the National Natural Science Foundation of China (NNSFC) under
grant No. 12075076 and the  Natural Science Foundation Project of Henan Province under grant No. 242300420253.

\end{document}